\documentclass[sigplan]{acmart}

\usepackage{nopageno}

\usepackage[normalem]{ulem}
\usepackage{amsmath}
\usepackage{algorithm}
\usepackage{algorithmicx}
\usepackage{graphicx}
\usepackage{color}
\usepackage{setspace}
\usepackage[dvipsnames]{xcolor}
\usepackage{mathtools}
\usepackage[noend]{algpseudocode}
\usepackage{soul}

\usepackage{comment}
\usepackage{fancyhdr}
\usepackage{hhline}
\usepackage{array}
\usepackage{xspace}
\usepackage{url}
\usepackage{caption}
\usepackage{subfig}
\usepackage{fancyvrb}
\usepackage{siunitx}
\usepackage{booktabs}
\usepackage{pifont}
\usepackage{flushend}
\usepackage[us,12hr]{datetime}
\usepackage{dblfloatfix}

\usepackage{multirow}

\sloppypar

\begin{document}

\newcommand\kasraa[1]{\noindent{\color{red} {\bf \fbox{Kasraa}} {\it#1}}}
\newcommand\TODO[1]{\noindent{\color{blue} {\bf \fbox{TODO}} {\it#1}}}
\newcommand\methodname[1]{\textsc{\mbox{#1}}}
\newcommand\appname[1]{\textsf{\mbox{#1}}}
\newcommand\componentname[1]{\textsl{\mbox{#1}}}
\newcommand{\pluseq}{\mathrel{+}=}
\newcommand\mehran[1]{\noindent{\color{blue} {\bf \fbox{Mehran}} {\it#1}}}


\title{Die-Stacked DRAM: Memory, Cache, or MemCache?}

\newcommand{\authspace}[0]{\hspace{4pt}}
\newcommand{\authfromaffil}[0]{\hspace{1pt}}
\newcommand{\affilspace}[0]{\hspace{15pt}}


\author{
Mohammad~Bakhshalipour\authfromaffil{}$^\ddagger{}$$^\S{}$
\authspace{}
HamidReza Zare\authfromaffil{}$^\ddagger{}$
\authspace{}
Pejman~Lotfi-Kamran\authfromaffil{}$^\S{}$
\authspace{}
Hamid~Sarbazi-Azad\authfromaffil{}$^\ddagger{}$$^\S{}$
}

\affiliation{\vspace{0.4cm}
$^\ddagger$\authfromaffil{}Department of Computer Engineering, Sharif University of Technology\\$^\S$\authfromaffil{}School of Computer Science, Institute for Research in Fundamental Sciences (IPM)
\vspace{0.4cm}}

\fancyhead{}

\begin{abstract}
Die-stacked DRAM is a promising solution for satisfying the ever-increasing memory bandwidth requirements of multi-core processors. Manufacturing technology has enabled stacking several gigabytes of DRAM modules on the active die, thereby providing orders of magnitude higher bandwidth as compared to the conventional DIMM-based DDR memories. Nevertheless, die-stacked DRAM, due to its limited capacity, cannot accommodate entire datasets of modern big-data applications. Therefore, prior proposals use it either as a sizable memory-side cache or as a part of the software-visible main memory. Cache designs can adapt themselves to the dynamic variations of applications but suffer from the tag storage/latency/bandwidth overhead. On the other hand, memory designs eliminate the need for tags, and hence, provide efficient access to data, but are unable to capture the dynamic behaviors of applications due to their static nature.

In this work, we make a case for using the die-stacked DRAM partly as main memory and partly as a cache. We observe that in modern big-data applications there are many hot pages with a large number of accesses. Based on this observation, we propose to use a portion of the die-stacked DRAM as main memory to host hot pages, enabling serving a significant number of the accesses from the high-bandwidth DRAM without the overhead of tag-checking, and manage the rest of the DRAM as a cache, for capturing the dynamic behavior of applications. In this proposal, a software procedure pre-processes the application and determines hot pages, then asks the OS to map them to the memory portion of the die-stacked DRAM. The cache portion of the die-stacked DRAM is managed by hardware, caching data allocated in the off-chip memory. Through detailed evaluations of various big-data applications, we show that our proposal improves system performance by 28\% and up to 139\% over the previous best-performing proposal.
\end{abstract}

\maketitle

\thispagestyle{empty}

\section{Introduction}

The increase in core count of chip multiprocessors (CMPs) has been driving the designs into the memory bandwidth wall, mainly because of pin count limitations~\cite{bakhshalipour2018fast, Huh:2001:EDS:645988.674164, rogers:scaling}. Current CMPs with tens of cores already lose performance because of the limited bandwidth of DIMM-based DDR memories, and the problem is exacerbated as the number of cores increases~\cite{kgil:picoserver, lotfi:sop}. Therefore, continuing performance scaling through core count scaling requires a commensurate enhancement in the bandwidth of the memory system.

Emerging die-stacked DRAM technology is a promising solution to fulfill the ever-increasing bandwidth requirements of multi-core processors. The progress in manufacturing has enabled stacking several DRAM modules on the active die using high-density through-silicon vias (TSVs). Compared to traditional DIMM-based DDR memories, die-stacked DRAM provides several orders of magnitude higher bandwidth, but with approximately the same latency~\cite{hmc2014hybrid, chang2013reevaluating, chou2017batman, Lee_SMA, sodani2015intel, sodani2016knights}. In recent years, several models have been developed for the die-stacked DRAM~\cite{wideio, standard2013high, micronhmc, hmc2014hybrid, 3dics}, and many commercial vendors have planned to use such models in their products~\cite{nvidia-volta, nvidia-pascal, amd-fiji, amdradeon,  lakka2012xilinx, sodani2015knights}.

\begin{figure}[t]
 \centering
 \includegraphics[width=.48\textwidth]{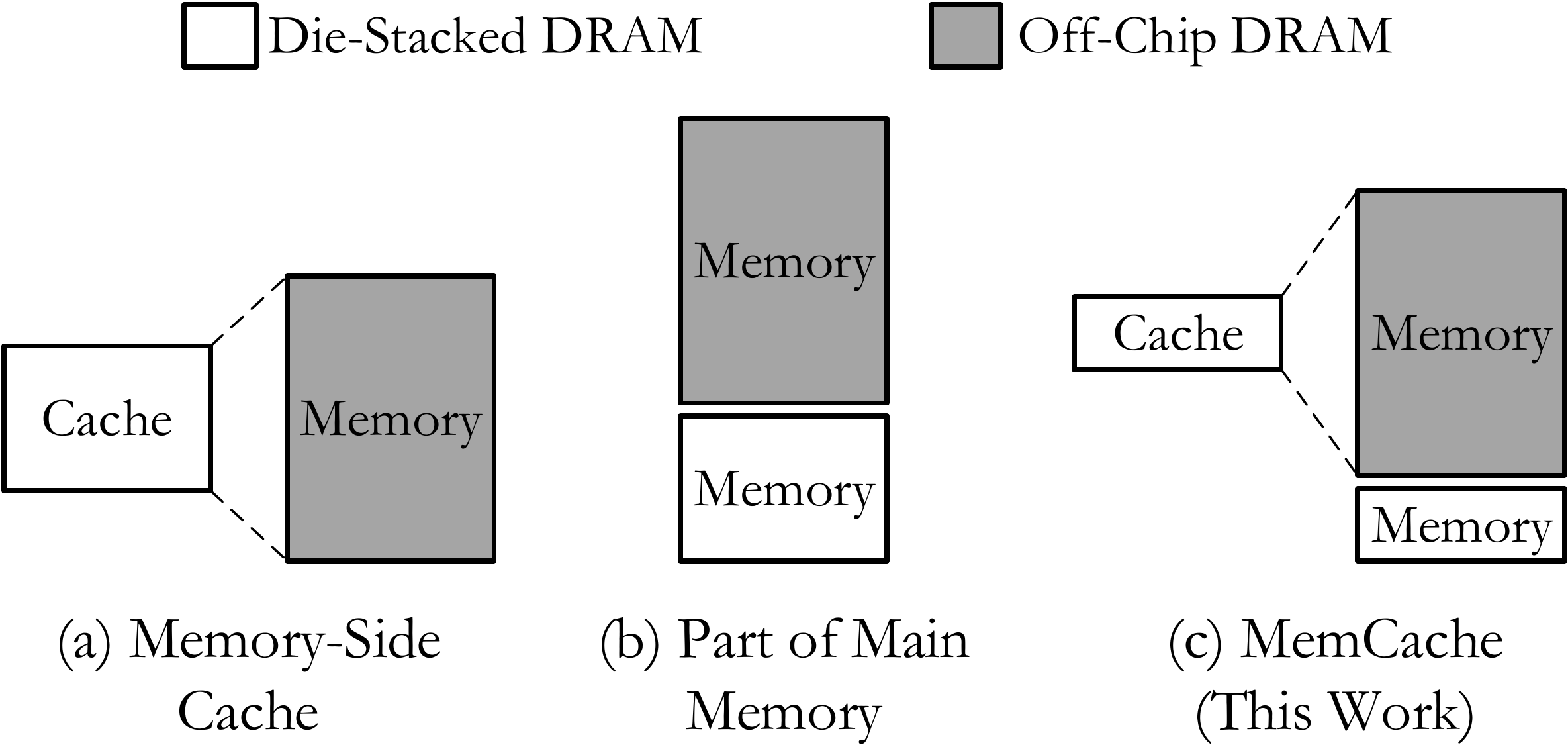}
 \caption{Comparison of proposals for the die-stacked DRAM.
 \label{fig:scheme_cache_mem_memcache}}
 
\end{figure}

One critical feature of the die-stacked DRAM is its limited capacity. Technological constraints, such as power-delivery and thermal limitations, restrict the size of the die-stacked DRAM to utmost a few gigabytes~\cite{healy2009study, lewis2011designing}. As such, it cannot accommodate the whole datasets of modern big-data applications with the datasets ranging from hundreds of gigabytes to a few terabytes. Consequently,  prior proposals use the die-stacked DRAM either as a memory-side cache~\cite{el2013dual, hameed2013simultaneously, jang2016efficient, jevdjic:unison, jevdjic:footprint, jiang2010chop, Lee:2015:FAT:2749469.2750383, Loh:2009:EED:1669112.1669139, loh:efficiently, qureshi:alloy, sim2012mostly, sim2013resilient, volos2017fat, yu2017banshee, zhao2007exploring} or as a \emph{part} of software-visible main memory~\cite{agarwal2015unlocking, Agarwal:2015:PPS:2694344.2694381, dong2010simple, loh2012challenges, meswani:heterogeneous, Sim:2014:THM:2742155.2742158}, as shown in Figure~\ref{fig:scheme_cache_mem_memcache}. 

DRAM caches can \emph{quickly react to changes} in the data working sets of applications, and use the DRAM capacity more effectively by evicting cold data and keeping actively-accessed objects. However, the DRAM cache designs \emph{suffer from the tag overhead}, which is substantially more than the tag cost of traditional SRAM caches. As the size of the die-stacked DRAM is quite large, managing it as a cache requires megabytes of storage, somewhere in the system, for storing the tag information. As placing such a large structure in the active die (i.e., SRAM) is impractical, most proposals put the tags in the die-stacked DRAM itself~\cite{Gulur:2014:BDC:2742155.2742160, huang2014atcache, jevdjic:unison, loh:efficiently, qureshi:alloy, yu2017banshee}. Nevertheless, placing the tags in the die-stacked DRAM can add significant latency to the critical path of cache accesses, as the DRAM should be accessed twice; once for the tag and then for the data. Some approaches~\cite{jevdjic:unison, qureshi:alloy} store the tags and data next to each other and stream them out together in a single access to avoid this serialization latency overhead. While these approaches are able to mitigate the latency overhead of tag-checking, they still incur significant extra bandwidth overhead.

On the other hand, using die-stacked DRAM as a part of main memory eliminates the main drawback of caches: \emph{there is no need for tags}. Consequently, no tag storage is required, and the tag latency/bandwidth overhead is eliminated. Memory, however, \emph{cannot respond to dynamic variations} in the data working sets of applications, due to its static nature. In memory designs, over time, there would be many pieces of cold data in the die-stacked DRAM that are not currently being used, wasting its capacity. Prior work~\cite{meswani:heterogeneous} proposed to swap pages between die-stacked and off-chip DRAM periodically, using run-time Operating System (OS) support. Upon each period, the OS identifies hot and cold pages in off-chip and die-stacked DRAM, respectively, and swaps them in order to have the currently-used pages in the high-bandwidth DRAM. Such approaches, nonetheless, are not cheaply-realizable and present significant challenges to both software and hardware. First, as OS intervention is very time-consuming, the periodic intervals should be large enough (e.g., hundreds of milliseconds) to amortize the enormous costs of interrupts, page migrations, and TLB shootdowns. As such, the pages that are highly-utilized for a short duration of time cannot be captured by these approaches. Second, as full OS pages should be transferred between die-stacked and off-chip DRAM, there is a significant bandwidth overhead associated with these approaches. Third, as the OS has no precise information about the utilization of pages at run-time, it cannot robustly rank them, and correctly identify hot and cold pages. 

In this work, we make a case for using the die-stacked DRAM partly as main memory and partly as a cache (Figure~\ref{fig:scheme_cache_mem_memcache}-c). We corroborate prior work~\cite{Chou:2014:CTM:2742155.2742157, dong2010simple, meswani2014toward} that using the whole capacity of the die-stacked DRAM as a part of the main memory is suboptimal to using it as a cache. However, we observe that, in modern big-data applications, there are numerous hot pages with a large number of accesses. Based on this observation, \emph{we classify application datasets into two distinct categories: hot datasets (hot pages), and transient datasets (transient pages)}. Hot datasets refer to data objects that are used \emph{steadily} by an application and \emph{serve a significant fraction of memory accesses} throughout the whole execution of an application. Transient datasets point to data structures that \emph{are utilized only for a short period of time}. In this work, we suggest \methodname{MemCache} for exploiting such heterogeneity in the access behavior of applications in the context of multi-gigabyte die-stacked DRAM. We suggest to use a portion of the die-stacked DRAM as a part of main memory and allocate hot pages in this part and use the rest of the capacity of the die-stacked DRAM as a hardware-managed cache to capture the transient datasets of applications. By allocating hot pages in the memory portion of the die-stacked DRAM, a considerable fraction of accesses are served from the high bandwidth memory without the overhead of tag-checking. The cache portion of the die-stacked DRAM remains intact and caches the transient datasets, providing quick responses to the dynamic variations in the datasets of applications.

To identify hot pages, we use a static profile-based approach before the execution of an application. A software procedure, incorporated into the compiler, pre-processes the application and sorts the pages based on their access frequency. Then it picks the top pages and asks the OS to map them to the memory portion of the die-stacked DRAM. We show that by using a representative input-set, such an offline analysis can classify pages robustly.

We evaluate \methodname{MemCache} against the state-of-the-art cache and memory proposals for the die-stacked DRAM and show that it significantly outperforms them on various big-data applications. Compared to a baseline without die-stacked DRAM, \methodname{MemCache} offers 114\% performance improvement on average and up to 309\%. Meanwhile, \methodname{MemCache} outperforms the best-performing prior design (\methodname{Banshee Cache}~\cite{yu2017banshee}) by 28\% on average and up to 139\%.

\section{Background}
\label{sec:background}

Modern big-data applications have vast datasets that dwarf capacity-limited SRAM caches and reside in memory~\cite{bakhshalipour2018domino, ferdman:clearing}. Such applications frequently access the off-chip memory for data, putting significant pressure on the DRAM modules. Consequently, substantial performance is lost solely because of bandwidth limitations of the off-chip memory. 

Recent research suggests using the die-stacked DRAM to break the bandwidth wall~\cite{Black_2006_DSM_1194816_1194860, Loh_3d}. As die-stacked DRAM cannot accommodate the whole datasets of big-data applications, prior proposals use it either as a large cache or as a part of main memory. In this section, we explore and discuss the design space of the die-stacked DRAM. 

\subsection {Die-Stacked DRAM as a Memory-Side Cache}
\label{sec:background:cache}

One major challenge in architecting a giga-scale DRAM cache is microarchitecting the tags. Since the size of die-stacked DRAMs is in the range of several gigabytes, maintaining the tag information of such a large structure requires megabytes of storage. For example, the tag array of a 4~GB block-based\footnote{By a block-based design, we refer to a die-stacked DRAM cache whose caching granularity is equal to the block size of the processor's SRAM caches (e.g., 64~bytes). In contrast, by a page-based design, we refer to a DRAM cache whose caching granularity is more coarse-grained (e.g., 4~KB).} DRAM cache requires in excess of 64~MB of storage. Managing the DRAM as a page-based cache reduces the tag storage to nearly 14~MB, which is still significant. Obviously, affording such large structures in SRAM is impractical; hence, most practical approaches store the information within the die-stacked DRAM itself. However, storing the tag information in the die-stacked DRAM imposes latency and bandwidth overheads, because, before every data access, the corresponding tag information should be checked via accessing the die-stacked DRAM. Many pieces of prior work targeted this overhead and suggested techniques to improve latency and/or bandwidth efficiency of cache accesses. 

Various design objectives (e.g., hit latency, bandwidth-efficiency and hit ratio) have been considered in architecting giga-scale DRAM cache designs. \methodname{Alloy Cache}~\cite{qureshi:alloy} is a block-based cache design that targets minimizing hit latency using a direct-mapped organization. \methodname{Alloy Cache} locates data and the corresponding tag, adjacently, and streams them out upon each access. This way, the latency of tag-checking is eliminated from the cache access if the access hits in the DRAM cache. Nevertheless, \methodname{Alloy Cache} suffers from two fundamental inefficiencies: (1) extra die-stacked DRAM bandwidth is consumed for accessing tag information, and (2) low hit ratio because of being a block-based cache and employing a direct-mapped organization.

\methodname{Unison Cache}~\cite{jevdjic:unison} improves the hit ratio of a DRAM cache by caching the data at the page granularity with a set-associative structure. As the size of the tag array, even for page-based designs, is significant, \methodname{Unison Cache} likewise co-locates the tag and data in the die-stacked DRAM. Caching the data at the page granularity improves the hit ratio but imposes substantial bandwidth overhead, because many blocks of a page are never used during the residency of the page in the DRAM cache~\cite{jang2016efficient, jevdjic:footprint}. Moreover, employing a set-associative organization necessitates searching all cache ways before touching the data, which adds latency and bandwidth overheads to the cache accesses.

Naively implementing a page-based design not only does not improve performance but may also harm it, mainly because of bandwidth inefficiency~\cite{jang2016efficient, jevdjic:footprint, jiang2010chop, yu2017banshee}. To reduce the traffic of page-based caches, \methodname{Unison Cache} takes advantage of a predictor for fetching the blocks within a page that will be used while the page is in the cache (i.e., referred to as the \emph{footprint} of the page~\cite{jevdjic:footprint}). Upon a cache miss, only those blocks within the requested page that are predicted to be used will be brought into the cache to reduce the traffic. While this technique (i.e., footprint caching) is useful at reducing the bandwidth usage, its efficiency is restricted by the accuracy of the predictor. For every misprediction, either the cache suffers from an extra miss due to not having the requested block in the cache, or non-useful blocks will be brought into the cache, wasting valuable bandwidth of off-chip and die-stacked DRAM. Moreover, the naive implementation of a set-associative structure requires two serialized accesses to the DRAM cache to get a cache block: (1) one access to read the tags and identify the location of the piece of data (i.e., way) and (2) another access to read the piece of data. To reduce the cache access latency, \methodname{Unison Cache} uses a \emph{way predictor} to access the requested block with two overlapped read operations for tag and data. Unfortunately, every time the predictor makes a mistake, the cache has to be accessed one more time, which increases the cache access latency and wastes valuable cache bandwidth. Moreover, even when the prediction is correct, extra bandwidth is consumed for accessing tag information, which puts further pressure on the die-stacked DRAM.

Another kind of DRAM caching approaches, referred to as \methodname{Tagless Dram Cache (TDC)}~\cite{jang2016efficient, Lee:2015:FAT:2749469.2750383}, align the granularity of caching with OS pages to track the tags of cache-resident pages within the Page Table and Translation Lookaside Buffers (TLBs). With such approaches, whenever a TLB miss occurs, the information of residency of the page in the DRAM cache (e.g., whether the page is cached or not and the location of the page in the DRAM cache if it is cached) are fetched together with the corresponding Page Table Entry (PTE). This way, the latency of tag-checking becomes virtually zero, but at the cost of significant complexities that are pushed into both software and hardware. In such approaches, the contents of the TLBs of \emph{all} the cores should (seem to) be coherent. Therefore, they use system-wide TLB shootdowns for updating the content of all TLBs whenever a piece of data is replaced in the DRAM cache. In addition to the complexities that are imposed because of run-time hardware-software co-operations, \emph{frequent} TLB shootdowns cause significant performance degradation due to costly software interventions and present a scalability challenge, as the latency of system-wide TLB shootdowns increases with increasing the core count~\cite{meswani:heterogeneous, romanescu2010unified}.

Another major drawback of these methods is the massive bandwidth overhead which is consumed because of fetching data at page granularity. Techniques like footprint caching~\cite{jevdjic:footprint} have limited applicability for such schemes, as these methods are restricted to use otherwise-unused bits in the PTEs to store footprint metadata, which is not adequately-spacious storage. For example, recent 64-bit Intel Xeon Phi processors~\cite{sodani2015knights} use 64-bit PTEs, in which, 18-bits are left unused. However, a 4~KB page includes 64 cache blocks, and hence, 64-bits are required to \emph{precisely} record the corresponding footprint information (i.e., one bit for each block, representing whether or not the block was touched during the last residency of the page in the cache). \methodname{Footprint-augmented TDC (F-TDC)}~\cite{jang2016efficient} attempts to solve this problem by storing $m$-line granularity footprint metadata, where a single bit is used to represent the existence of $m$ cache blocks in the footprint of each page, in order to fit within the capacity limitations of PTEs. Unfortunately, this strategy leads to the over-fetching of data and bandwidth pollution, as the precise knowledge of the residency of cache blocks is lost. On each misprediction, up to $m$ blocks are over-fetched, which leads to further bandwidth pollution. The problem is even more significant because modern big-data applications heavily use large (e.g., 4~MB) and huge (e.g., 1~GB) OS pages for increasing the TLB reach, as discussed in the recent work~\cite{papadopoulou2015prediction}. Employing $m$-line granularity footprint prediction for 1~GB pages requires saving 1 bit for every 450~K blocks. As a consequence, upon every misprediction, up to 450~K cache blocks are over-fetched, imposing an unbearable bandwidth overhead.

Lastly, since these methods use \emph{different} address spaces for caches (i.e., SRAM caches plus a DRAM cache) and off-chip memory, and because they do not flush the \emph{stale} entries of SRAM caches after each page remapping, they sacrifice consistency among physical addresses and are prone to result in wrong execution\footnote{We do not further discuss this problem, as it is entirely addressed in the recent work~\cite{yu2017banshee}, where it is called the \emph{address consistency problem}.}. 


Recent work, namely the \methodname{Banshee Cache}~\cite{yu2017banshee}, addresses the first problem by \emph{coalescing} Page-Table updates. The \methodname{Banshee Cache} caches the details of recently-remapped pages (i.e., pages that were recently placed in or evicted from the DRAM cache) in an auxiliary SRAM structure, named the \componentname{Tag Buffer}. Whenever the occupancy of the \componentname{Tag Buffer} exceeds a certain threshold, a software interrupt is triggered to flush its entries into the Page Table and the TLBs. By provisioning enough capacity for the \componentname{Tag Buffer} (e.g., $\sim$5~KB), the frequency of Page Table updates decreases, which helps amortize the high cost of software interrupts. However, employing such a structure imposes SRAM storage overhead and may present scalability challenges as the core count increases. With more cores, the pressure on the \componentname{Tag Buffer} increases as the frequency of page re-mapping increases, resulting in more frequent flushes and costly software interventions. Therefore, the size of the \componentname{Tag Buffer} should commensurately be increased when increasing the number of cores, to preserve performance scalability. Furthermore, the complexity of run-time hardware-software co-operation is still a problem, as software is responsible for flushing the \componentname{Tag Buffer} into the Page Table and TLBs. To reduce the over-fetching problem, the \methodname{Banshee Cache} uses a bandwidth-aware replacement policy, in which, pages are replaced \emph{lazily} to lower the bandwidth pressure on die-stacked and off-chip DRAM. While the replacement policy is effective at reducing the movement of pages, it still consumes extra bandwidth to access page metadata (e.g., tag and replacement information) stored in the die-stacked DRAM. Finally, the \methodname{Banshee Cache} solves the address consistency problem of previously-proposed Page Table based approaches by using the \emph{same} address space for the die-stacked and off-chip DRAMs.

\subsection {Die-Stacked DRAM as a Part of Main Memory}

Using the die-stacked DRAM as a part of main memory eliminates the necessity of tags, and hence tag overheads, the major drawback of cache designs. The OS allocates some of the pages in the die-stacked DRAM physical address space and the rest in the off-chip DRAM. Such a scheme, however, cannot respond to dynamic changes in the data working sets of applications. Even if the OS applies intelligent algorithms for allocating hot pages in the high-bandwidth die-stacked memory, over time, there would be many pages in the die-stacked DRAM that are not currently being used, wasting its precious capacity. This happens because modern big-data applications have many \emph{dynamic data-dependent behaviors}, which cannot be detected/exploited statically before the execution of an application at page-allocation time~\cite{Huang_DIP, Sudan_MID}. 

Several pieces of prior work~\cite{dong2010simple, meswani:heterogeneous, Sim:2014:THM:2742155.2742158} propose \emph{swapping} pages between the high- and low-bandwidth memory, periodically, to capture the dynamic data-dependent behavior of applications throughout their execution. In the state-of-the-art approach~\cite{meswani:heterogeneous}, regularly, the OS \emph{ranks} all pages based on their usage and moves hot pages into the die-stacked DRAM and cold pages out. 

Unfortunately, such approaches suffer significantly from the massive cost of page swapping concerning both latency and bandwidth. In each interval and after the page swapping, the OS should update \emph{all} PTEs and shoot down \emph{all} TLBs for coherence, which takes a considerable amount of time. Moreover, exchanging many pages between die-stacked and off-chip DRAM consumes considerable bandwidth, as whole pages (and not their footprints, as in cache designs) should be transferred. For example, swapping two 4~KB pages, \componentname{P1} and \componentname{P2}, requires 16~KB of data transfer bandwidth: 4~KB for reading \componentname{P1}, 4~KB for writing \componentname{P1}, 4~KB for reading \componentname{P2}, and 4~KB for writing \componentname{P2}. If an application has many transient pages, the number of pages that should be swapped in each interval is also increased, exacerbating the bandwidth inefficiency of these approaches. Therefore, these methods perform the page swapping at \emph{very coarse granularity} (e.g., hundreds of milliseconds) to amortize the associated latency and bandwidth overheads. Consequently, the pages that are highly-utilized for a short duration of time (which we call \emph{transient pages}) cannot be captured by these approaches. 

Another drawback of these methods is the limited ability to detect hot pages. As the OS has no precise information about the utilization of pages at \emph{run-time}, it cannot accurately rank them, and correctly identify hot and cold pages. Usually, there is a single bit in each PTE that indicates whether the page is used or not, and no knowledge of access frequency is available at run-time. Therefore, precisely ranking the pages at execution time requires non-trivial changes to both hardware and software, which makes such designs more complex.

\section{Motivation}

We first show that, in modern big-data applications, there is a discrepancy in page usage: some of the pages are used more frequently than the others. Then we compare two extreme use cases of the die-stacked DRAM: (1) whole die-stacked DRAM as a cache, and (2) whole die-stacked DRAM as a part of main memory. Finally, we motivate to use a part of the die-stacked DRAM as main memory and the rest as a cache.

\subsection{Hot Pages}
\label{motiv:hot-pages}
Corroborating many pieces of prior work (e.g.,~\cite{agarwal2015unlocking, Agarwal:2015:PPS:2694344.2694381, jiang2010chop, meswani2014toward, Sudan_MID, volos2016effective, volos2017fat, yu2017banshee}), we observe that, in modern big-data applications, there are many hot pages with a large number of accesses. Hot pages refer to the phenomenon that some pages are accessed with higher frequency than the others. This causes a small fraction of an application's memory working set to serve a significant fraction of accesses (e.g., 10\% of data working set serve 80\% of requests).

\begin{figure}[b]
 \centering
 \includegraphics[width=.48\textwidth]{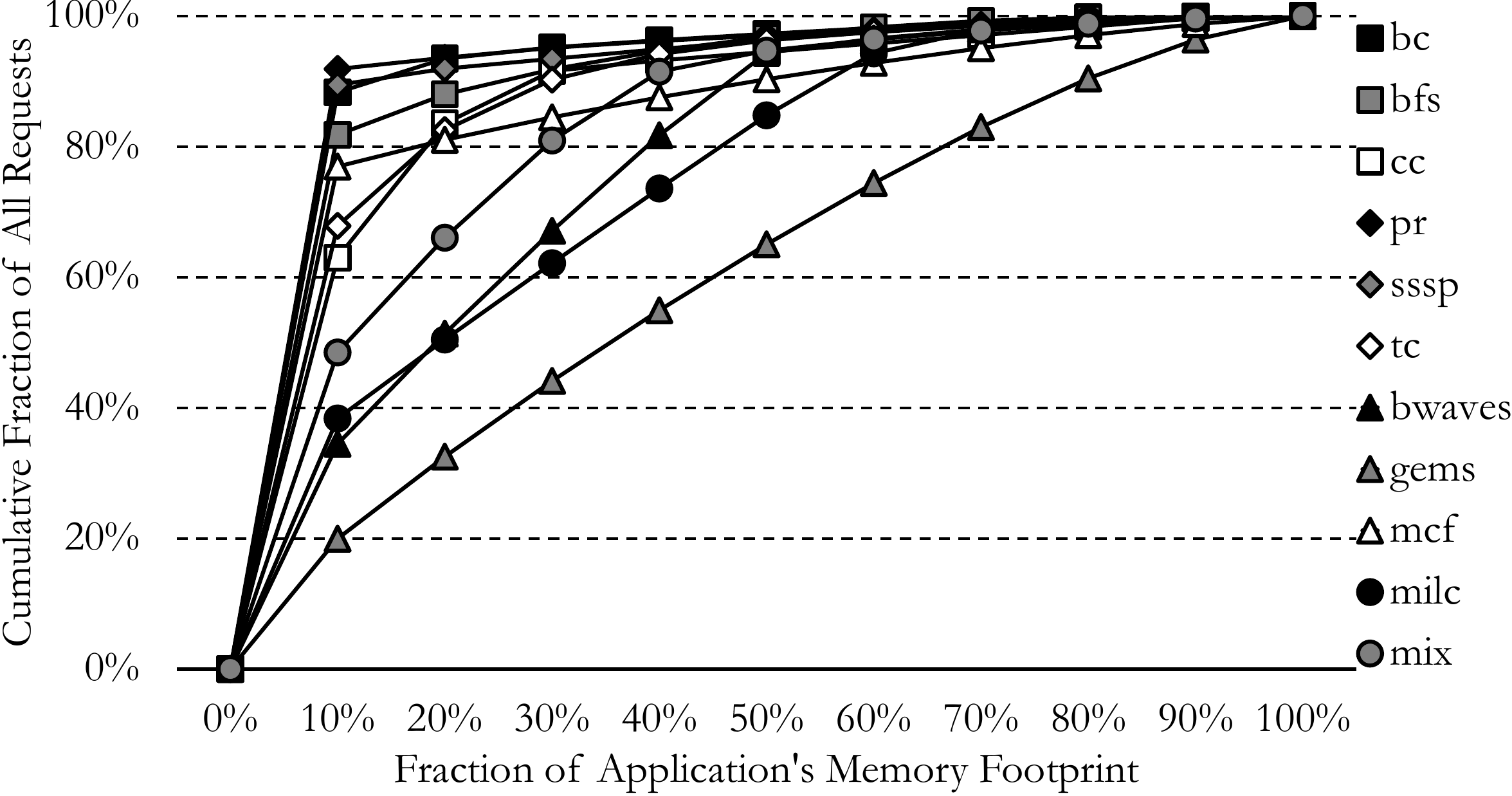}
 \caption{Cumulative distribution of data accesses with pages sorted from hottest to coldest.
 \label{fig:hot_pages}}
\end{figure}

Figure~\ref{fig:hot_pages} shows the cumulative distribution of accesses to the main memory for several big-data applications. The figure shows that most of the applications contain many hot pages that serve a significant fraction of total memory accesses. For example, in \appname{pr}, only 10\% of pages ($\sim$7~GB of memory footprint) serve more than 91\% of requests. The figure further indicates that applications consist of many transient pages. As an example, in \appname{pr}, the remaining 9\% of accesses are distributed among 90\% of pages with nearly the same access frequencies. \emph{We find that hot pages that are responsible for a significant fraction of accesses are used {steadily} throughout the whole execution of applications}\footnote{We elaborate more on this observation later in Section~\ref{eval:how_long}.}. In contrast, transient pages are used only for a short period of time.

The existence of hot pages motivates identifying and allocating them into the die-stacked memory to efficiently exploit the higher bandwidth that it provides. However, if we use the whole capacity of the die-stacked DRAM as a part of main memory and allocate hot pages into it, transient pages will be served from the off-chip DRAM. In contrast, if we use the die-stacked DRAM as a cache, we can serve both hot and transient datasets from the die-stacked DRAM, but have to pay the tag overhead (cf.~Section~\ref{sec:background:cache}). 

\subsection{Full-Cache Versus Full-Memory}
\label{motiv:cache-vs-mem}
\begin{figure}[t]
 \centering
 \includegraphics[width=.48\textwidth]{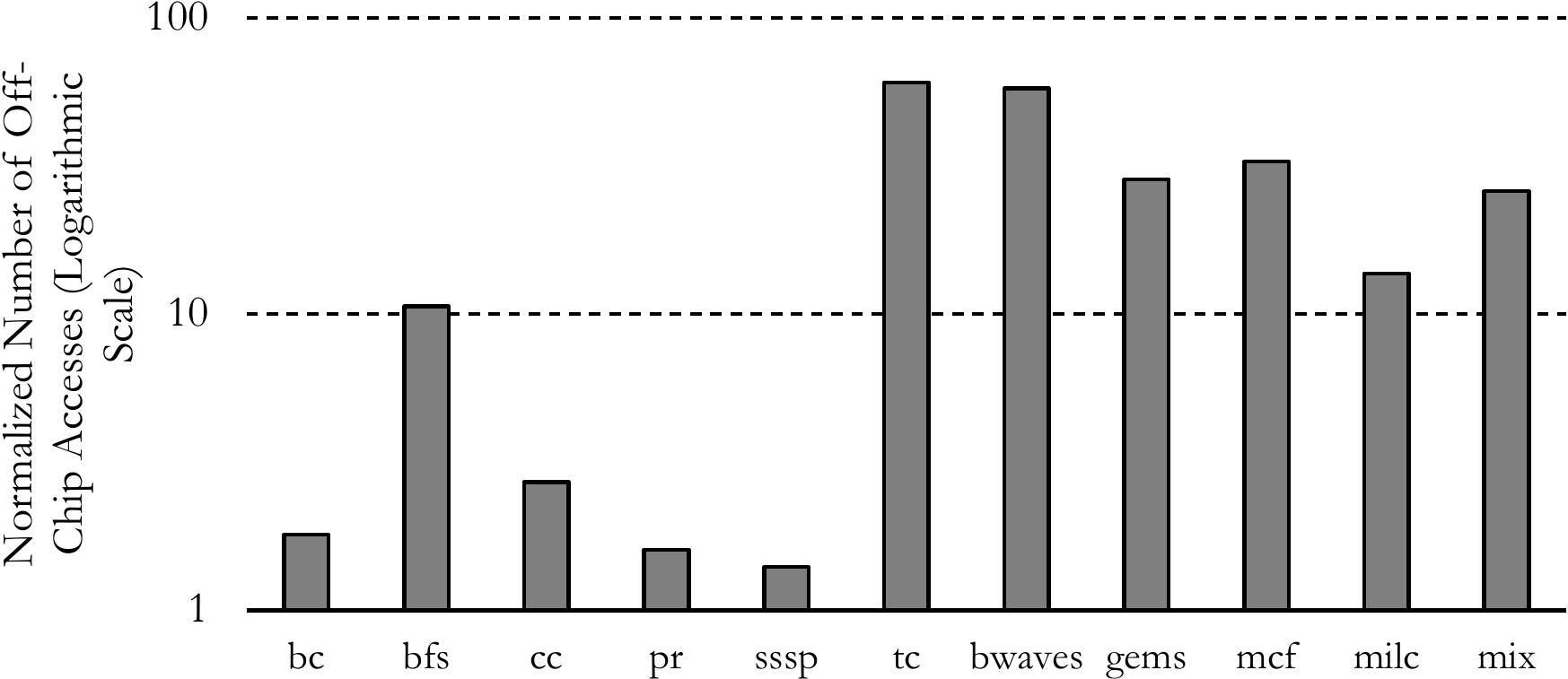}
 \caption{The number of off-chip accesses of a 4~GB die-stacked DRAM memory as compared to that of a cache.
 \label{fig:mem-vs-cache}}
\end{figure}

If the goal is to use the die-stacked DRAM as a part of main memory, the best one can do is to place pages with the highest number of references (i.e., hottest pages) in the die-stacked memory. Figure~\ref{fig:mem-vs-cache} shows the number of requests that go off-chip in a 4~GB die-stacked memory normalized to a 4~GB page-based cache for several big-data applications. In this experiment, we go over the sequence of accesses and find pages with the highest number of accesses and place them in the die-stacked memory.

As Figure~\ref{fig:mem-vs-cache} clearly shows, using the whole capacity of the die-stacked DRAM as a memory considerably increases the number of accesses to the bandwidth-limited off-chip DRAM, as compared to using it as a cache. It comes from the fact that a die-stacked DRAM as memory cannot adapt itself to the dynamic run-time changes in the data working set of applications and yet is not large enough to capture the entire datasets of big-data applications. The results indicate that even an ideal memory, which is oracularly filled with the hottest pages, increases the number of off-chip accesses from $1.4\times$ in \appname{sssp} to $60.5\times$ in \appname{tc}, as compared to a cache design. The results are consistent with those of prior work~\cite{ Chou:2014:CTM:2742155.2742157, dong2010simple, meswani2014toward} concluded that, for big-data applications, memory designs are inferior to caches at reducing the number of off-chip accesses.

\subsection{What Fraction of Die-Stacked DRAM Can Be Turned into Memory?}
\label{sec:motiv:what_frac}
Even though using the whole capacity of the die-stacked DRAM as a part of main memory increases the number of off-chip accesses as compared to caches, the existence of hot pages with a considerable number of accesses may suggest that we can use a part of the die-stacked DRAM as a memory for hosting a subset of such hot pages and use the rest of the capacity as a cache for capturing the dynamic data-dependent behavior of applications. 

The single major drawback of using the die-stacked DRAM as memory is its inability to respond to dynamic changes of applications. Over time, many transient pages will emerge in the data working set of an application, and since such pages were not detected and allocated in the die-stacked DRAM, a large number of requests will be sent to the bandwidth-limited off-chip memory. This causes memory designs to offer an order of magnitude lower hit\footnote{In this paper, by \emph{hit}, we refer to a state where the requested data is in the die-stacked DRAM, regardless of the fact that it is managed as a cache or as a part of memory. Analogously, by \emph{miss}, we refer to the state, in which, the requested data is not in the die-stacked DRAM.} ratio (higher off-chip accesses) as compared to cache designs (cf.~Section~\ref{motiv:cache-vs-mem}). Therefore, the mixture design (i.e., the design which uses the die-stacked DRAM partly as main memory and partly as cache) \emph{will lose nothing compared to the full-cache design, if it is able to offer a hit ratio as high as the cache}. 

In order to determine what fraction of a die-stacked DRAM can be turned into memory without losing the benefits of the cache design, we first consider the die-stacked DRAM as a set of \emph{frames}. Each frame in the die-stacked DRAM is a \emph{physical location} where a piece of data at the granularity of a page reside, regardless of the fact that the die-stacked DRAM is managed as a cache or as a memory. We first consider the whole die-stacked DRAM as a sizeable page-based cache and calculate the \componentname{Average number of Hits per Frame (AHF)}. Then \emph{we turn each frame into memory and allocate the hottest unallocated page in it until the \emph{\componentname{AHF}} becomes smaller than that of the cache}. Algorithm~\ref{alg:ahf} summarizes the steps that we take in order to determine the fraction of die-stacked DRAM that can be turned into a memory.

\begin{algorithm}[t]
\caption{Calculate the Memory Fraction}\label{alg:ahf}
\begin{algorithmic}[1]
\footnotesize
\State $cacheAHF=totalCacheHits/dramFrames;$
\State $pages=sort(pages);$\Comment{{\scriptsize Sort pages based on their access count}}
\State $memFrames=0;$
\State $totalAccessesToMemFrames=0;$\newline
\While{$(totalAccessesToMemFrames\geq memFrames\times cacheAHF)$}
\State $hotPage = pages.top();$\Comment{{\scriptsize The hottest unallocated page}}
\State $memFrames\mathrel{++};$\Comment{{\scriptsize Allocate the hot page in the die-stacked memory}}
\State $pages.removeTop();$
\State $totalAccessesToMemFrames\pluseq hotPage.accessCount;$
\EndWhile\label{euclidendwhile}
\State \textbf{done}\newline
\State $memFraction=memFrames/dramFrames;$

\end{algorithmic}
\end{algorithm}

\begin{figure}[b]
 \centering
 \includegraphics[width=.48\textwidth]{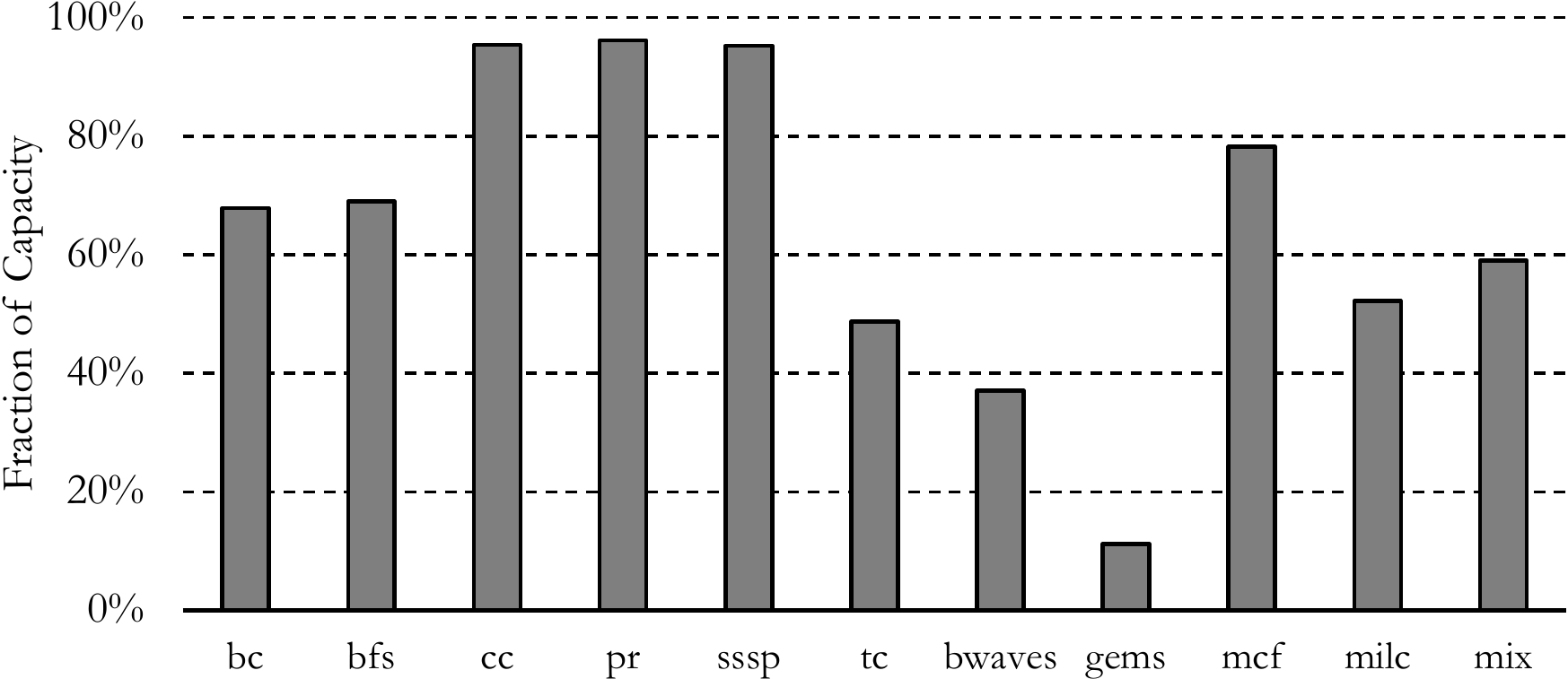}
 \caption{The fraction of the die-stacked DRAM that can be managed like memory and still offers the same hit ratio as a page-based cache.
 \label{fig:max_memory}}
\end{figure}

Figure~\ref{fig:max_memory} shows the fraction of a 4~GB die-stacked DRAM that can be managed as memory and still offer the same hit ratio as a 4~GB page-based cache. As the figure shows, a significant fraction of the die-stacked DRAM can be managed as a memory without negatively increasing the number of off-chip accesses. The fraction ranges from 11\% in \appname{gems}\footnote{\appname{gems} is an exceptional case, in which, requests are distributed among all pages with nearly the same frequency (i.e., it does not have a considerable number of hot pages. cf.~Figure~\ref{fig:hot_pages}).} to 96\% in \appname{pr}, with an average of 65\%. This means that, on average, 65\% of a 4~GB DRAM can be turned into memory without losing the benefits of a full-cache design. Note that, this is the fraction, at which, the \emph{hit ratio} of the design that uses the die-stacked DRAM partly as main memory and partly as cache remains intact as compared to the full-cache design. While the hit ratio is important for die-stacked DRAM organizations, prior work showed that minimizing \emph{hit latency}~\cite{qureshi:alloy} or efficiently utilizing the \emph{die-stacked DRAM bandwidth}~\cite{yu2017banshee} are of equal importance if not more. As memory organizations generally offer lower access latency than cache organizations and do not waste DRAM bandwidth for replacements/tag manipulations, we expect that even a larger fraction of the die-stacked DRAM ($> 65\%$ on average) can be managed as a part of main memory. We determine the fraction of the die-stacked DRAM's capacity that can be managed as a memory using a performance sensitivity analysis in Section~\ref{sec:eval:capacity_partition}.

Taking advantage of this observation, we propose to use a portion of the die-stacked DRAM as a part of the main memory and allocate hot pages to it and use the rest of the capacity of the die-stacked DRAM as a hardware-managed cache to capture the transient data-dependent behavior of applications. By allocating hot pages in the memory portion of the die-stacked DRAM, a significant fraction of accesses is served from the high-bandwidth memory without the overhead of tag-checking (bandwidth and latency). The cache portion of the die-stacked DRAM also remains intact and caches the transient pages, providing quick responses to dynamic variations in the datasets of applications.

\section{The Proposal}

In order to exploit the heterogeneity in the access behavior of applications to hot and transient pieces of data, we propose \methodname{MemCache}, a design that uses a part of the die-stacked DRAM as main memory and the rest as a cache. The software is responsible for filling the die-stacked memory with the identified hot pages. The transient pages are also served from the hardware-managed cache portion of the die-stacked DRAM. For every \mbox{Last-Level SRAM Cache (LLSC)} miss, if the miss address is mapped to the memory portion of the die-stacked DRAM, it is served by the die-stacked memory (i.e., without the overhead of tag-checking). Otherwise, it checks the DRAM cache tag array. In case of a cache hit, the request is served by the cache portion of the die-stacked DRAM, and in case of a cache miss, it proceeds to the off-chip memory.

\subsection{Memory Portion}
\label{sec:proposal:memory_portion}
The memory frames of the die-stacked DRAM should be filled with the hot pages of applications. First off, the hot pages of an application should be identified. We use a \emph{profiling} approach in order to distinguish hot pages from cold ones. A software procedure, incorporated into the compiler, using profiling, determines which \emph{virtual pages} of the application are hot. To do so, the software procedure sorts virtual pages based on their access count, in descending order, and then, conveys this information as clues to the page allocation unit of the OS. 

There are quite a few ways to implement the profiling step needed in our mechanism, two of which are more common and are elaborated upon here. One approach is that the compiler profiles the program by simulating the behavior of the on-chip cache hierarchy of the target machine~\cite{ebrahimi2009techniques}. The simulation is used to gather the access counts of pages, which are then used to identify hot pages. Note that, such a profiling approach does not necessitate an accurate timing simulation of the processor and cache hierarchy; rather, it requires simply a trace-driven simulation of memory operations in order to gather information about accesses that are not captured in on-chip SRAM caches (i.e., LLSC misses and evictions). Another approach for implementing the profiling step is relying on hardware support. In this strategy, the target machine provides support for profiling operations (e.g., \emph{Informing Load Operations}~\cite{Horowitz_IMO}). With this support, throughout the profiling run, the compiler gathers the access counts of application's pages, and at the end, sorts them accordingly. In this paper, we use the first approach as it requires no change in the hardware. 

\begin{figure}[t]
 \centering
 \includegraphics[width=.48\textwidth]{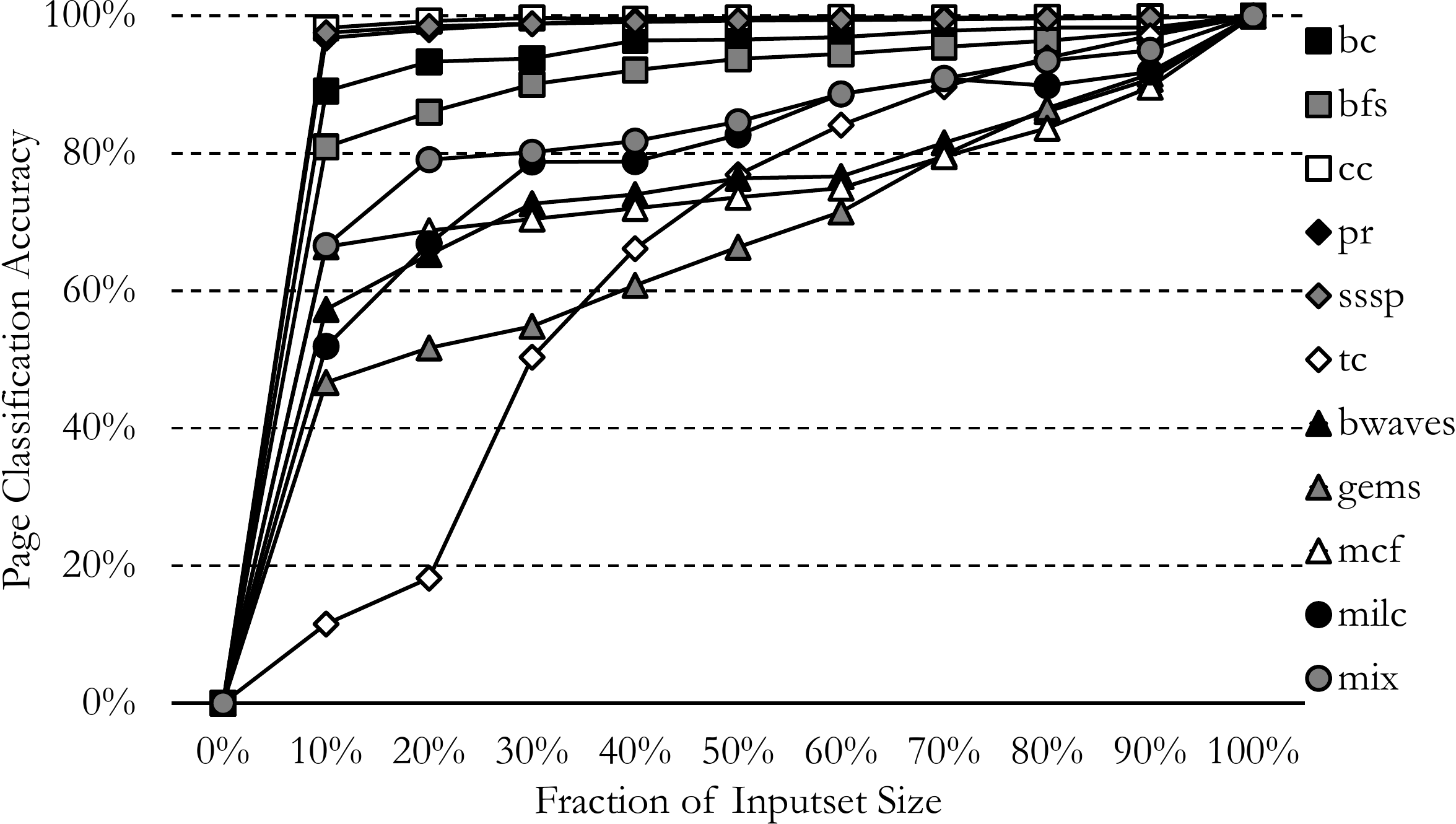}
 \caption{The effect of sample inputset size on the accuracy of page classification.
 \label{fig:classification-accuracy}}
\end{figure}

Identifying hot pages by the compiler demands for a \emph{sample inputset} on which the profiling step should be run. The sample inputset must be large enough to enable the compiler to observe various pages, gather adequate access counts on them, and distinguish hot pages from cold ones by sorting them. To shed light on how large the inputset should be, we perform the following experiment: First, we use quite large inputsets\footnote{The details of applications and their inputsets are explained in Section~\ref{sec:methodology}.}, such that simulating the on-chip cache hierarchy on them, produces 10~B LLSC misses/evictions for each application. Based on these inputsets, we find the hottest pages of every application that fill a 4~GB memory. We refer to the set of these hot pages (identified by evaluating 10~B LLSC misses/evictions) as \emph{ideal pages}. Then, we redo the same experiment based on just a part of the same inputsets, that results in 1~B LLSC misses/evictions for every application, and again identify the hottest pages that fill a 4~GB memory. Finally, we compare these hot pages with ideal pages and report the ratio of similar pages to all pages as the \emph{accuracy of page classification}. We also repeat the same experiment based on 2~B, 3~B, \ldots{}, and 9~B LLSC misses/evictions. Figure~\ref{fig:classification-accuracy} shows how the accuracy of page classification varies with increasing the sample inputset size. 

A critical observation we make here is that \emph{by observing a small fraction of the inputset, a static profile-based approach is able to classify pages accurately}.  This is especially true for modern throughput-oriented big-data graph applications. For example, in \appname{cc}, classifying pages into hot and cold based on observing the first 1~B accesses yields 98\% of the accuracy of an ideal classifier which is aware of the future and classifies pages based on the whole 10~B accesses\footnote{Later in Section~\ref{eval:how_long}, we show that not only is such a page classification accurate, but it also produces robust results. That is, the pages that are classified as hot remain hot throughout the whole execution of an application. }. On average across all applications, by observing only 1~B accesses, a profile-based approach is able to classify pages accurately, realizing roughly three-fourths of the accuracy of an ideal mechanism. Note that this fairly simple mechanism (i.e., profile-based classification of pages) works well mainly because of the fact that the size of the die-stacked DRAM is quite large. That is, the die-stacked DRAM is capacious enough to accommodate thousands and even millions of pages, and a few misclassifications do not have a significant effect on the overall accuracy\footnote{Generally, the accuracy of probabilistic methods (such as the employed profile-based method that estimates a set of hot pages for the whole execution of an application based on observing a small fraction of accesses) for \emph{Top-N Recommendation} increases with increasing \emph{N}~\cite{Barbieri_APM}.}.

After detection of hot pages, their details are coded into the program binary. Whenever the program gets executed, the \componentname{Loader} passes the required information of hot pages to the OS. Then, the OS tries\footnote{Sometimes, there might be conflicting preferences/constraints that get in the way of the OS allocating a page to a specific physical address space~\cite{bovet2005understanding}. Resolving such issues is beyond the scope of this paper. } to map such hot pages to physical locations that belong to the memory portion of the die-stacked DRAM. For the OS, allocating pages in the die-stacked and off-chip memory is similar to the same operations in Non-Uniform Memory Architecture (NUMA)~\cite{Larowe_ECM} systems\footnote{Here, NUMA regions are further broken into a die-stacked memory region and an off-chip memory region.}. Therefore, the OS, as well as system libraries, is able to use the same memory allocation algorithms for allocating pages in the die-stacked and off-chip memory. Furthermore, other operations, like virtual address translation and virtual memory management, for the die-stacked and off-chip memory are quite the same as in a NUMA system.

\subsection{Cache Portion}
The cache portion of the die-stacked DRAM filters requests to the pages that are allocated in the off-chip memory (i.e., pages that have not been identified as hot). The cache is transparent to the software and can be managed using any DRAM cache management technique. In this paper, we evaluate three cache architectures (\methodname{Alloy Cache}~\cite{qureshi:alloy}, \methodname{Unison Cache}~\cite{jevdjic:unison}, and \methodname{Banshee Cache}~\cite{yu2017banshee}) for the cache portion of the \methodname{MemCache}.

\subsection{Memory-Cache Capacity Partitioning}
The optimal capacity of the die-stacked DRAM which can be turned into a part of the main memory directly depends on the applications. For applications with significant hot pages (e.g., \appname{pr}), it is better to dedicate less space for the cache and more to memory. On the other hand, for applications with few hot pages (e.g., \appname{gems}), partitioning the capacity towards more cache would be beneficial. However, dynamically changing the partition based on the running applications makes the system more complex, especially when the system is supposed to run multiple applications and \emph{frequently} switch from one application to another. 

In this paper, we consider two variants for \methodname{MemCache}; one design which its partitioning is tuned based on the application, and another which has fixed partitioning. In the first design, named \methodname{MemCache-D}, the compiler determines what fraction of the die-stacked DRAM capacity should be devoted to the memory and what fraction to the cache. To do so, the compiler runs Algorithm~\ref{alg:ahf}, thereby finding a fairly suitable partitioning\footnote{As discussed in Section~\ref{sec:motiv:what_frac}, in Algorithm~\ref{alg:ahf}, first a full-cache design is simulated. Then, the frames of the die-stacked DRAM are gradually turned into memory until the whole design becomes susceptible to fall behind a full-cache design.}. Then, the OS, using a specific instruction, announces the partitioning to the memory controllers. This way, the memory controllers become able to redirect requests to their correct locations\footnote{By checking the physical address of a request and comparing it with the physical address of the last frame which has been devoted to memory, the memory controller decides whether to send the request to the memory portion of the die-stacked DRAM or serve it from the cache portion.}. \methodname{MemCache-D} adapts itself based on the application behavior and is supposed to offer higher performance than a design that its partition is determined statically. However, it needs to change the partitioning for every application, which obviously is a costly operation, and may not be effective when the system runs multiple applications, frequently switching from one to another\footnote{When the memory-cache partitioning of the die-stacked DRAM changes, not only is a system-wide TLB shootdown required, but also the valid data in the memory frames and modified data in the cache frames should be written back to the off-chip memory if the functionality of the frames changes.}. Therefore, we believe that \methodname{MemCache-D} is a suitable design for systems where a specific application (or a set of specific applications\footnote{In this paper, we evaluate this technique when multiple \emph{dissimilar} applications co-run on the CMP.}) runs for a long time (e.g., datacenter applications). 

The other variant of \methodname{MemCache} we consider in this work is \methodname{MemCache-S}. In \methodname{MemCache-S}, a \emph{fixed} fraction of the die-stacked DRAM is devoted to the memory and the rest to the cache. The partitioning is determined by the user, \emph{at boot time}, and is then reported to the OS by the BIOS, along with other information. While a fixed partitioning might not give the best performance for all of the applications, it realizes a significant fraction of the performance benefit of dynamic application-based partitioning (i.e., \methodname{MemCache-D}). In this paper, we choose the capacity partitioning of \methodname{MemCache-S} based on the performance sensitivity analysis of all applications (Section~\ref{sec:eval:capacity_partition}) and show that such a static design offers a level of performance close to that of a design that picks the best memory-cache partitioning for each application.

The case where multiple applications share the processor, and the system frequently switches from one application to another, presents complications to \methodname{MemCache} and other hybrid memory systems where multiple dissimilar memories are used (e.g., \componentname{DRAM+NVM} systems where \componentname{DRAM} has characteristics entirely different from \componentname{NVM}, or even \componentname{NUMA} systems where accessing \componentname{Local Memory} has lower latency than \componentname{Remote Memories}). The problem arises from the fact that when the hot pages of an application (or set of applications) are being allocated in the die-stacked memory, information about the hot pages of other applications that will possibly be running in the future is unknown; therefore, filling the die-stacked DRAM merely based on the current applications may result in suboptimal performance if future applications will be more bandwidth-hungry and require more capacity from the die-stacked DRAM. Various proposals (e.g., \cite{Agarwal_TAP, knyaginin2018profess, li2017utility, Ogasawara_NMM, Verghese_OSS}) have suggested to optimize memory management in such situations typically by gradually or periodically migrating application pages between different types of memories based on factors like programming model, application's criticality, sharing degree, and so on. Such approaches are orthogonal to \methodname{MemCache} and can augment it to provide performance/fairness benefits; however, we leave the analysis of enhancing our method with these techniques for future work, promoting a clearer understanding of our contribution in the context of prior die-stacked DRAM literature.

\section{Methodology}
\label{sec:methodology}

We use \componentname{ZSim}~\cite{sanchez-zsim} to simulate a system whose configuration is shown in Table~\ref{table:config}. We model the system based on one \componentname{Sub-NUMA Cluster (SNC)} of Intel's \componentname{Knights Landing} processor~\cite{sodani2016knightsieeemicro}. The chip has 16 OoO cores with an 8~MB LLSC and a 4~GB die-stacked DRAM. One channel is used for accessing off-chip DRAM, providing a maximum bandwidth of 21~GB/s, and four channels are responsible for establishing communications with the die-stacked DRAM, providing up to 84~GB/s bandwidth. In comparison, Intel's \componentname{Knights Landing} processor has four \componentname{SNCs}, offering 72 cores, 36~MB LLSC (512~KB per core; 8~MB per 16 cores), 16~GB die-stacked DRAM (4~GB per \componentname{SNCs}), 90~GB/s off-chip peak bandwidth (1.25~GB/s per core; 20~GB/s per 16 cores), and up to 400~GB/s stacked bandwidth (5.5~GB/s per core; 88~GB/s per 16 cores). DRAM modules are modeled based on \componentname{DDR3-1333} technology, parametrized with data borrowed from commercial device specifications~\cite{ddr3}. Physical addresses are mapped to memory controllers at 4~KB granularity. The link width among memory controllers and die-stacked DRAM is 16~B, but the minimum data transfer size is 32~B~\cite{yu2015imp, yu2017banshee}.

\begin{table}[h]
\sffamily
 \begin{center}
  \caption{Evaluation parameters.}
  \label{table:config}
   { 
    \resizebox{.48\textwidth}{!}{
        \begin{tabular}{| c || c |}
         \hline
          {\bf {Parameter}}      & {\bf {Value}}  \\
         \hline
         \hline
          {Cores}                     & Sixteen 4-wide OoO cores, 2.8~GHz \\
         \hline
          {L1-D/I}                    & 32~KB, 2-way, 1-cycle load-to-use \\
         \hline
          {L2 Cache}                  & 8~MB, 16-way, 15-cycle hit latency \\
         \hline
          {Die-Stacked DRAM}          & 4~GB, 4 channels, 4~KB interleaved \\
         \hline
          {Off-Chip DRAM}             & 1 channel, up to 21~GB/s bandwidth \\
         \hline
          \multirow{3}{*}{DRAM Modules} & DDR 1333~MHz, 8~KB row buffer \\
                                      & 4 ranks/channel, 8 banks/rank, 16-byte bus width \\
                                      & tCAS-tRCD-tRP-tRAS = 10-10-10-24 \\
         \hline
        \end{tabular}
    }
   } \small
 \end{center}
\end{table}

\subsection{Workloads}
Table~\ref{table:workloads} summarizes the key characteristics of our simulated workloads. We use all graph processing workloads from \componentname{GAPBS}~\cite{beamer2015gap} and run them with the \componentname{Twitter} inputset~\cite{kwak2010twitter}. While the primary target of this work (and the systems that employ die-stacked DRAM) is throughput-oriented workloads like graph processing, yet for reference, we also include workloads from the \componentname{SPEC}~\cite{henning2006spec} benchmark suite. We choose four \componentname{SPEC} benchmarks whose memory footprints exceed 4~GB and run them with the \componentname{reference} inputset in \componentname{RATE} mode (i.e., all sixteen cores run the same single-thread program). We also consider a \appname{mix} of the four \componentname{SPEC} programs in which the processor executes four copies of each program.


\begin{table}[t]
\sffamily
\scriptsize
 \begin{center}
  \caption{Application parameters.}
  \label{table:workloads}
    \resizebox{.42\textwidth}{!}{
        \renewcommand{\arraystretch}{1}
        \begin{tabular}{| c || c | c |}
         \hline
          {\bf {Application}} & {\bf {LLSC MPKI}}  & {\bf {Memory Footprint (GB)}} \\
         \hline
         \hline
           \multicolumn{3}{|c|}{\bf {GAPBS}} \\
           \hline
         \hline
           {\appname{bc}}     & 61.4  & 92.9 \\
         \hline
           {\appname{bfs}}    & 32.9  & 113.3 \\
         \hline
           {\appname{cc}}     & 85.6  & 9.6 \\
         \hline
           {\appname{pr}}     & 129.9 & 76.5 \\
         \hline
           {\appname{sssp}}   & 73.4  & 140.1 \\
         \hline
           {\appname{tc}}     & 12.5  & 53.4 \\
         \hline
         \hline
           \multicolumn{3}{|c|}{\bf {SPEC}} \\
           \hline
         \hline
           {\appname{bwaves}} & 17.6  & 13.1 \\
         \hline
           {\appname{gems}}   & 26.7  & 12.6 \\
         \hline
           {\appname{mcf}}    & 66.9  & 26.1 \\
         \hline
           {\appname{milc}}   & 17.0  & 10.2 \\
         \hline
           {\appname{mix}}   & 17.8  & 15.4 \\
         \hline
        \end{tabular}
    }
 \end{center}
\end{table}

\subsection{Die-Stacked DRAM Organizations}

We compare the following die-stacked DRAM organizations:\\

\begin{figure*}[t]
 \centering
 \makebox[\textwidth][c]{\includegraphics[width=\textwidth]{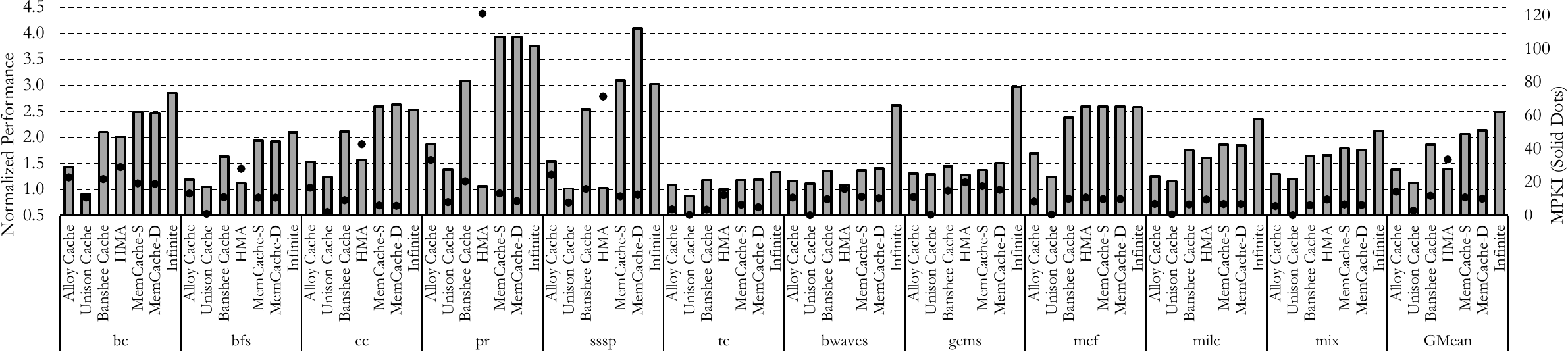}}
 \caption{Performance improvement (bars) and MPKI (solid dots) of evaluated designs.
 \label{fig:performance}}
\end{figure*}

\noindent
\textbf{\methodname{Alloy Cache}:} A state-of-the-art block-based cache design that uses a direct-mapped organization to minimize hit latency. \methodname{Alloy Cache} has a simple structure and was shown to be quite effective~\cite{qureshi:alloy}. \\

\noindent
\textbf{\methodname{Unison Cache}:} A state-of-the-art page-based four-way set-associative cache design with LRU replacement policy that uses a footprint predictor~\cite{jevdjic:footprint} to reduce the off-chip bandwidth pressure. Moreover, it uses a way predictor to avoid the serialization latency of tag and data accesses. In this paper, in order to decouple the results of cache design from the accuracy of predictors, we use \emph{perfect} predictors for fetching expected-to-use blocks and the cache way where the data is located at. For modeling a perfect footprint predictor, we follow a similar methodology as prior work~\cite{yu2017banshee}. We first profile applications offline and measure the average number of blocks that are used from cache pages (i.e., average footprint size). In timing simulations, whenever we want to bring a page into the die-stacked DRAM, we transfer as many cache blocks as the average footprint size of the application. However, unlike prior work~\cite{yu2017banshee}, we consider an accurate 1-line granularity footprint predictor (instead of 4-line granularity).\\

\noindent
\textbf{\methodname{Banshee Cache}:} A state-of-the-art page-based four-way set-associative cache design that tracks the tags of cache-resident pages in the Page Table and \componentname{Tag Buffer}. We consider a 1K-entry 8-way associative \componentname{Tag Buffer} (5.5~KB of SRAM storage) and set the baseline sampling coefficient to 10\% based on the original proposal~\cite{yu2017banshee}. Whenever the occupancy of the \componentname{Tag Buffer} exceeds 70\%, it should be flushed to the Page Table (and TLBs) to make all information coherent. We consider 25{$\mu{}s$} latency for the software procedure that updates the Page Table and shoots down TLBs based on the original proposal.\\

\noindent
\textbf{\methodname{HMA}:} The whole die-stacked DRAM is used as a part of main memory. Upon every $10^8$ LLSC misses, a software procedure swaps the pages between the die-stacked and off-chip DRAM to have currently-used pages in the high-bandwidth memory. For every page in the system, we consider a counter that counts the number of accesses to that page. Upon each OS interval, we \emph{precisely} sort the pages and allocate highly-used ones in the high bandwidth memory. 

By accurately modeling the bandwidth of page swapping, we observed that the performance (IPC) of this technique consistently drops to \textit{zero} in all workloads because of the massive bandwidth overhead. The simulated workloads already utilize the bandwidth of both die-stacked and off-chip DRAM, and hence, there is not much unused bandwidth left for swapping numerous large pages on each interval. Increasing the page swapping period (e.g., once every 10 seconds) may amortize the associated bandwidth overhead, but doing so will further reduce the ability of this technique to respond to the dynamic changes in the memory access behavior of applications (i.e., transient datasets). In this paper, we model \emph{an idealized case}, in which, page swapping consumes \emph{no bandwidth} and happens instantaneously with \emph{no latency}.\\

\noindent
\textbf{\methodname{MemCache-S}:} Based on the performance sensitivity analysis (see Section~\ref{sec:eval:capacity_partition}), 3~GB of the die-stacked DRAM is used as a part of main memory, and the remaining 1~GB operates as a cache. Software allocates the identified hot pages in the memory portion of the die-stacked DRAM based on the \emph{warm-up} instructions (i.e., the instructions and the part of inputset which are used for identifying hot pages and filling the die-stacked memory are not used/counted in the actual experiments). Without loss of generality, we consider a \methodname{Banshee Cache} architecture for the cache portion of the die-stacked DRAM\footnote{In Section~\ref{sec:eval:cache_design}, we evaluate other cache architectures for the cache portion of \methodname{MemCache}, as well.}.\\

\noindent
\textbf{\methodname{MemCache-D}:} For every application, the memory-cache capacity partitioning is determined using Algorithm~\ref{alg:ahf}. The other specifications are the same as \methodname{MemCache-S}.\\

\noindent
\textbf{\methodname{Infinite}:} Die-stacked DRAM has infinite size and is used as a memory to host the \emph{whole} datasets of applications.

\subsection{Simulation Parameters}

For trace-driven experiments, we gather 16~billion LLSC misses, using the first 2~billion for warm-up and the rest for measurements. For timing analysis, we run the simulations for 200~billion instructions and use the first 20~billion for warm-up and the next 180~billion for measurements. We ensure that after the warm-up period, the die-stacked DRAM has become well-filled (i.e., no empty frame has remained), and the statistics are in the steady state. 

\section{Evaluation}

\label{sec:eval}

\subsection{Performance}

\label{sec:eval:perf}

Figure~\ref{fig:performance} shows the performance of all designs, normalized to a system without die-stacked DRAM. Moreover, solid dots in the figure indicate Misses Per Kilo Instructions (MPKI) of the competing designs. On average, \methodname{MemCache-S/D} offer 107\%/114\% performance improvement, which is 21\%/28\% higher than \methodname{Banshee Cache}, the best-performing prior proposal.

\methodname{Unison Cache} improves the performance by 12\% on average and has the lowest performance enhancement among the evaluated designs. While it offers a high hit ratio (thanks to the page-based organization and ideal footprint predictor), it suffers from inefficiencies in utilizing the die-stacked DRAM bandwidth, as we show in Section~\ref{sec:eval:on-chip-traffic}. Due to frequent replacements of large pages and checking/updating the tags, a significant bandwidth overhead is imposed on the die-stacked DRAM, preventing it from offering considerable performance improvement.

As compared to \methodname{Unison Cache}, \methodname{Alloy Cache} has better performance improvement (38\% on average), mainly because of imposing less bandwidth pressure on the die-stacked DRAM. \methodname{Alloy Cache} caches data at block granularity, and hence, consumes less bandwidth upon replacements. Moreover, it uses a direct-mapped organization which frees it from the tag-updating bandwidth (e.g., LRU bits). However, \methodname{Alloy Cache} still consumes significant bandwidth for probing the tag and speculatively loading data, as we show in Section~\ref{sec:eval:on-chip-traffic}. For minimizing hit latency, \methodname{Alloy Cache} co-locates tag and data adjacently and streams them out in a single access. Whenever a request hits in the DRAM cache, the latency of accessing data is a single DRAM access, but extra bandwidth is consumed for loading the tag. In case of a cache miss, not only the tag but also the speculatively-loaded data impose bandwidth overhead. Another drawback of \methodname{Alloy Cache} is its relatively high MPKI. Caching data at block granularity and employing a direct-mapped organization synergistically decrease the cache hit ratio because of not exploiting the spatial locality and the increased conflict misses.

\methodname{HMA} has no tag overhead but suffers extensively from the low hit ratio of the die-stacked DRAM. As the periods of page swapping (OS intervals) are too long, the technique is unable to serve requests to transient datasets. Therefore, requests to data objects that are highly utilized for only a short period of time are served from the bandwidth-limited off-chip memory. We emphasize that reducing the period of OS intervals is not feasible because of the colossal associated latency and the bandwidth overhead of swapping numerous large pages.

\methodname{Banshee Cache} tracks tags of cache-resident pages through TLBs and hence eliminates the latency of tag-checking. Moreover, its bandwidth-aware replacement policy significantly reduces the frequency of replacements, resulting in less bandwidth pressure on the die-stacked and off-chip DRAM. However, its performance is far from that of \methodname{Infinite}. Lazily replacing pages performed by the bandwidth-aware replacement policy of \methodname{Banshee Cache} reduces the hit ratio, resulting in serving more requests from the bandwidth-limited off-chip memory. Moreover, in order to reach the metadata of pages (e.g., tag and counter) that are stored in the die-stacked DRAM, \methodname{Banshee Cache} consumes considerable bandwidth, which prevents it from reaching the peak performance. Finally, costly software interrupts that read and flush the Tag Buffer entries to the Page Table and TLBs take significant system cycles (each accounts for tens of kilo-cycles), resulting in performance degradation. 

The performance improvement of \methodname{MemCache-S/D} ranges from 18\%/19\%  to 294\%/310\% with an average of 107\%/114\%. Compared to \methodname{Banshee Cache}, the best-performing previous proposal, \methodname{MemCache-S/D} improves the performance by 21\%/28\% on average and up to 85\%/138\%. The performance improvement is more evident in bandwidth-hungry throughput-oriented applications (e.g.,~\appname{pr}). The performance improvement of \methodname{MemCache-S/D} over \methodname{Banshee Cache} comes from: 

\begin{figure*}[t]
 \centering
 \makebox[\textwidth][c]{\includegraphics[width=\textwidth]{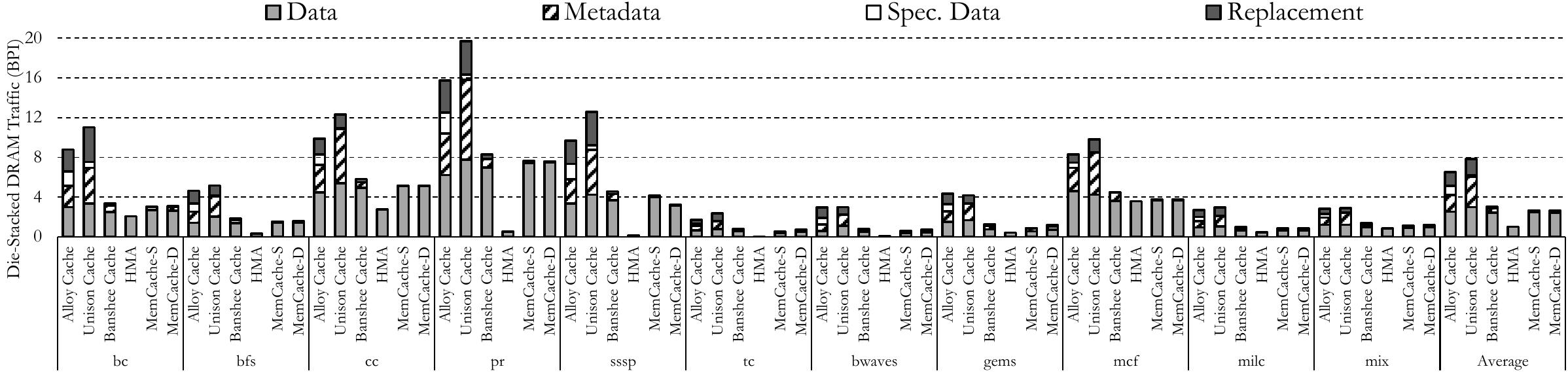}}
 \caption{The breakdown of die-stacked DRAM traffic.
 \label{fig:die-stacked-bandwidth}}
\end{figure*}

\begin{figure*}[b]
 \centering
 \makebox[\textwidth][c]{\includegraphics[width=\textwidth]{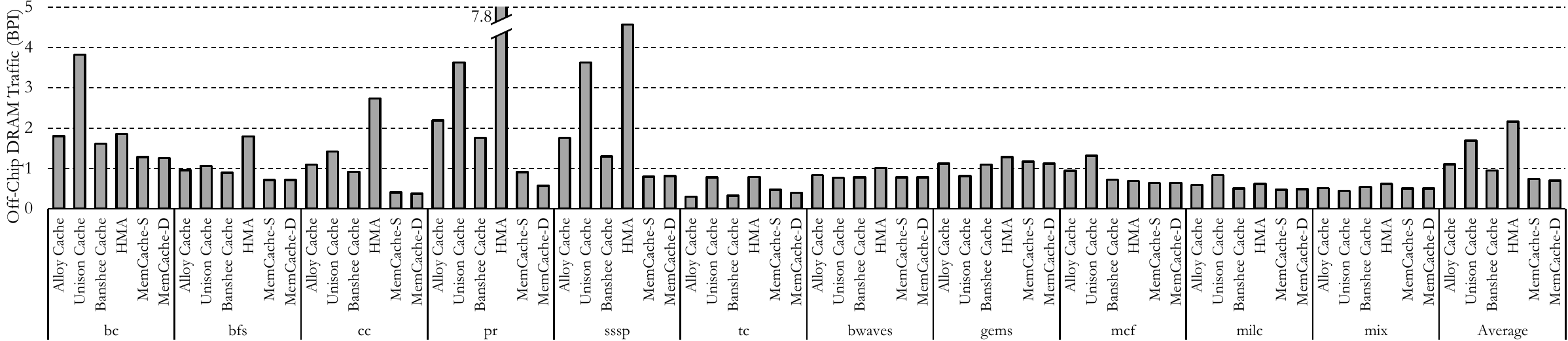}}
 \caption{Off-chip DRAM traffic of evaluated designs.
 \label{fig:off-chip-bandwidth}}
\end{figure*}

\begin{enumerate}
 \item \methodname{MemCache-S/D} mitigates wrong replacement decisions made by \methodname{Banshee Cache} (i.e., not caching pages\footnote{\methodname{Banshee Cache} caches a page only if, intuitively, the page is identified hotter than other pages resident on the same set. This is problematic when more than four hot pages are mapped to the same set, preventing having all/many of them in the high-bandwidth DRAM. Increasing the associativity may mitigate this problem, but will exacerbate the bandwidth pressure on the die-stacked DRAM as more bytes need to be be transferred upon each metadata access. \methodname{MemCache} virtually eliminates this problem by allocating hot pages in the memory portion of the die-stacked DRAM.}). Therefore, it reduces both miss ratio and off-chip bandwidth of \methodname{Banshee Cache}. On average, \methodname{MemCache-S/D} reduces the MPKI of \methodname{Banshee Cache} by 8\%/15\% and up to 36\%/45\%. Moreover, it reduces the off-chip bandwidth by 22\%/27\%, on average, and up to 56\%/59\%. 

 \item \methodname{MemCache} reduces the bandwidth pressure of the die-stacked DRAM that is imposed because of accessing metadata in \methodname{Banshee Cache} for either replacement or tag-probing. It comes from the fact that, a significant fraction of requests is served from the memory portion of the die-stacked DRAM, for which, no metadata bandwidth is consumed. On average, \methodname{MemCache-S/D} consumes 13\%/15\% less stacked bandwidth (up to 31\%/30\%) as compared to \methodname{Banshee Cache}.

 \item The number of costly software interrupts is reduced with \methodname{MemCache}, as compared to \methodname{Banshee Cache}. The main reason is that, with serving a significant fraction of requests from the memory portion of the die-stacked DRAM, there is a lighter load on the \componentname{Tag Buffer}, and hence, it is filled up more slowly. \methodname{MemCache-S/D} reduces the number of \componentname{Tag Buffer} flushes by 78\%/87\% on average and up to 94\%/99\%. 
 
\end{enumerate}

\methodname{MemCache-D} offers higher performance as compared to \methodname{MemCache-S}. \methodname{MemCache-D} outperforms \methodname{MemCache-S} by 7\% on average and up to 84\%, mainly because of more intelligent partitioning of the die-stacked DRAM capacity. However, in few cases, the performance improvement of \methodname{MemCache-D} is slightly less (up to 2\%) than that of \methodname{MemCache-S}. It is because we partition the die-stacked DRAM capacity (i.e., Algorithm~\ref{alg:ahf}) towards memory, \emph{conservatively}. That is, we dedicate frames to the memory as much as the whole design (i.e., the composition of memory and cache) becomes susceptible to fall behind the full-cache design. However, as discussed in Section~\ref{motiv:cache-vs-mem}, since the memory has better latency/bandwidth characterizations as compared to the cache, the optimal point in capacity partitioning might be where the memory fraction is slightly higher than what Algorithm~\ref{alg:ahf} determines. One solution may be enhancing the offline partitioning algorithm by considering factors other than hit ratio, such as bandwidth and/or latency. However, doing so adds complications to the offline process since the compiler should run a \emph{timing simulation} rather than simply a trace-driven simulation. We conclude that the minor performance improvement does not justify this increased complexity. 

Impractical \methodname{Infinite} has the highest average performance improvement. \methodname{Infinite} improves the performance by 149\% on average and up to 275\%. However, in several cases (e.g., \appname{cc}, \appname{pr}, \appname{sssp}, and \appname{mcf}), its performance is less than that of \methodname{MemCache-S/D} (in \appname{mcf} also less than \methodname{HMA}). This is because, with \methodname{Infinite}, applications benefit only from the bandwidth of the die-stacked DRAM and not the off-chip DRAM. Therefore, with \methodname{Infinite}, the total bandwidth available to applications is less than other approaches in which the bandwidth of both die-stacked and off-chip DRAM are available.

\begin{figure*}[b]
 \centering
 \makebox[\textwidth][c]{\includegraphics[width=\textwidth]{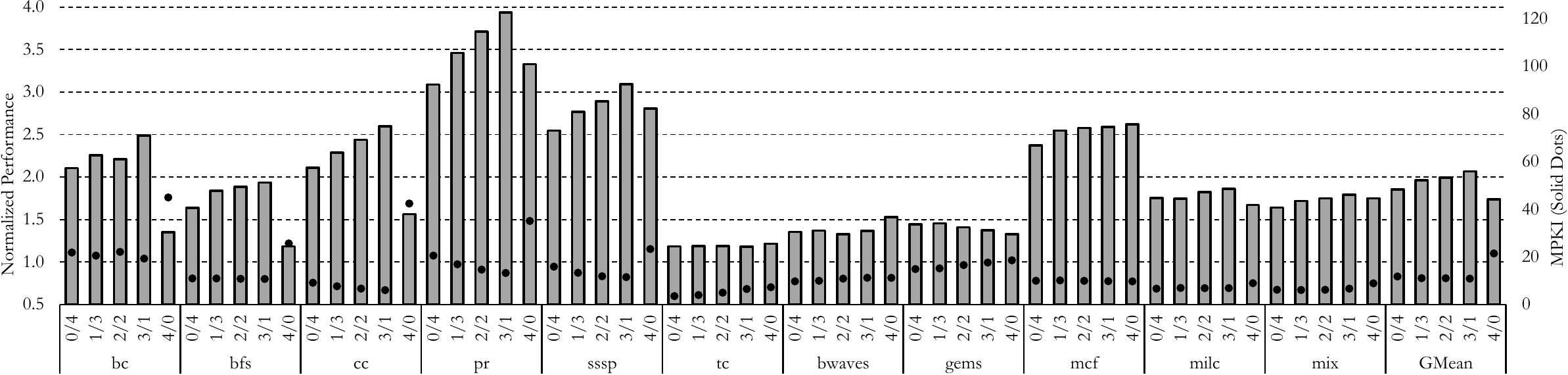}}
 \caption{The sensitivity of performance (bars) and MPKI (solid dots) to memory-cache capacity partitioning. Performance is normalized to a baseline without die-stacked DRAM. \textit{X/Y} represents a design that dedicates \textit{X}~GB to memory and \textit{Y}~GB to cache. 
 \label{fig:memcache-partition}}
\end{figure*}

\subsection{Die-Stacked DRAM Traffic}
\label{sec:eval:on-chip-traffic}

As throughput-oriented applications (like graph processing) are able to utilize the bandwidth of die-stacked DRAM, managing die-stacked DRAM in a bandwidth-efficient manner is crucial for reaching the peak performance. 

Figure~\ref{fig:die-stacked-bandwidth} shows the breakdown of die-stacked DRAM traffic for the competing designs. We use \emph{Bytes Per Instruction}~\cite{yu2017banshee} as the metric for bandwidth intensity of applications. The breakdown consists of \componentname{Data}, \componentname{Metadata}, \componentname{Spec. Data}, and \componentname{Replacement}. \componentname{Data} refers to the traffic that is used for reading and updating data in the die-stacked DRAM. \componentname{Metadata} includes the bandwidth consumed for reading/updating tag and replacement metadata (e.g., LRU bits). \componentname{Spec. Data} corresponds to the traffic that is imposed to speculatively load data for requests that miss in the die-stacked DRAM. This type of traffic is specific to \methodname{Alloy Cache} and \methodname{Unison Cache}. Finally, \componentname{Replacement} shows the bandwidth consumed for replacing cache blocks/pages. 

\methodname{HMA} has the lowest traffic as it touches the die-stacked DRAM only for data\footnote{Repeatedly, we do not model the bandwidth of page-swapping in \methodname{HMA}.}. \methodname{Alloy Cache} and \methodname{Unison Cache} impose an order of magnitude higher traffic to the die-stacked DRAM, mainly because of metadata accesses. The metadata traffic is higher in \methodname{Unison Cache} because it employs a set-associative structure and is required to update metadata information upon each access. \methodname{Banshee Cache} has lower bandwidth pressure because it has fewer metadata accesses and lazily replaces pages in the die-stacked DRAM. Meanwhile, it has higher bandwidth usage as compared to \methodname{MemCache}. By serving a significant fraction of requests from the memory portion of the die-stacked DRAM, \methodname{MemCache-S/D} reduces the die-stacked bandwidth usage of \methodname{Banshee Cache} by 13\%/15\%. The reduction mainly comes from the cut down on metadata and replacement traffic.

\subsection{Off-Chip DRAM Traffic}

Figure~\ref{fig:off-chip-bandwidth} shows the traffic of the off-chip DRAM for the evaluated designs. Off-chip DRAM traffic is more important than that of the die-stacked DRAM because more gap between the bandwidth of die-stacked and off-chip DRAM is expected with the future technologies~\cite{tran2016era}.

\methodname{HMA} has the highest traffic because of its lower hit ratio. As discussed previously, \methodname{HMA} is unable to capture the dynamic behavior of applications, due to its long OS intervals. Therefore, in this technique, virtually all requests to transient data objects are served from the off-chip DRAM, imposing a significant bandwidth overhead. \methodname{Unison Cache} and \methodname{Alloy Cache} impose traffic on the off-chip DRAM due to cache misses and dirty writebacks. \methodname{Banshee Cache} has a less off-chip bandwidth, as compared to \methodname{Unison Cache} and \methodname{Alloy Cache}, because it has fewer replacements. \methodname{MemCache} has the lowest off-chip bandwidth among the evaluated designs. The bandwidth reduction of \methodname{MemCache-S/D} over \methodname{Banshee Cache} is 22\%/27\% (up to 56\%/59\%) and comes from mitigating wrong replacement decisions made by \methodname{Banshee Cache}, resulting in higher hit ratio of the die-stacked DRAM.

\subsection{Sensitivity to Memory-Cache Capacity Partitioning}
\label{sec:eval:capacity_partition}

The optimal capacity partitioning between the cache and memory in the die-stacked DRAM depends on the applications. Figure~\ref{fig:memcache-partition} shows the performance and MPKI of \methodname{MemCache-S} as the fraction of memory varies. Applications that have many hot pages (e.g., \appname{pr}) favor dedicating smaller capacity to the cache and larger to the memory. On the other hand, for applications with significant transient pages (e.g., \appname{gems}), partitioning the capacity towards more cache is beneficial. Among the different partitionings, the design which devotes 3~GB to memory and 1~GB to cache offers the highest average performance; consequently, we choose this partitioning for the \methodname{MemCache-S}. We note that on average, cache design (i.e., \emph{0/4}) performs better than full-memory design (i.e., \emph{4/0}). This confirms that due to the mismatch between the dataset size and die-stacked DRAM capacity, full-memory is suboptimal to full-cache designs. However, in few cases and despite the fact that full-memory design has higher MPKI than the cache design, its performance is slightly better. This is due to the fact that a full-memory design completely eliminates the whole overhead associated with the evaluated cache scheme (i.e., \methodname{Banshee Cache}): no metadata overhead, no replacement, and no expensive software interrupt. 

An interesting data point is that, for applications that largely benefit from the die-stacked DRAM technology (i.e., the performance gets more than doubled), it is beneficial to dedicate more capacity to memory and less to cache. This is the main reason why the design that uses three-fourths of the die-stacked DRAM as memory offers a significant performance improvement of the dynamically-partitioned design. Modern big-data throughput-oriented applications (e.g., most of the graph-processing applications in our suite) have a significant number of hot pages (cf.~Section~\ref{motiv:hot-pages}) which can be hosted in the memory portion of the die-stacked DRAM, enabling better usage of the die-stacked DRAM technology, and hence higher performance improvement. 

\subsection{The Architecture of the Cache Portion}
\label{sec:eval:cache_design}

\begin{figure}[t]
 \centering
 \includegraphics[width=.48\textwidth]{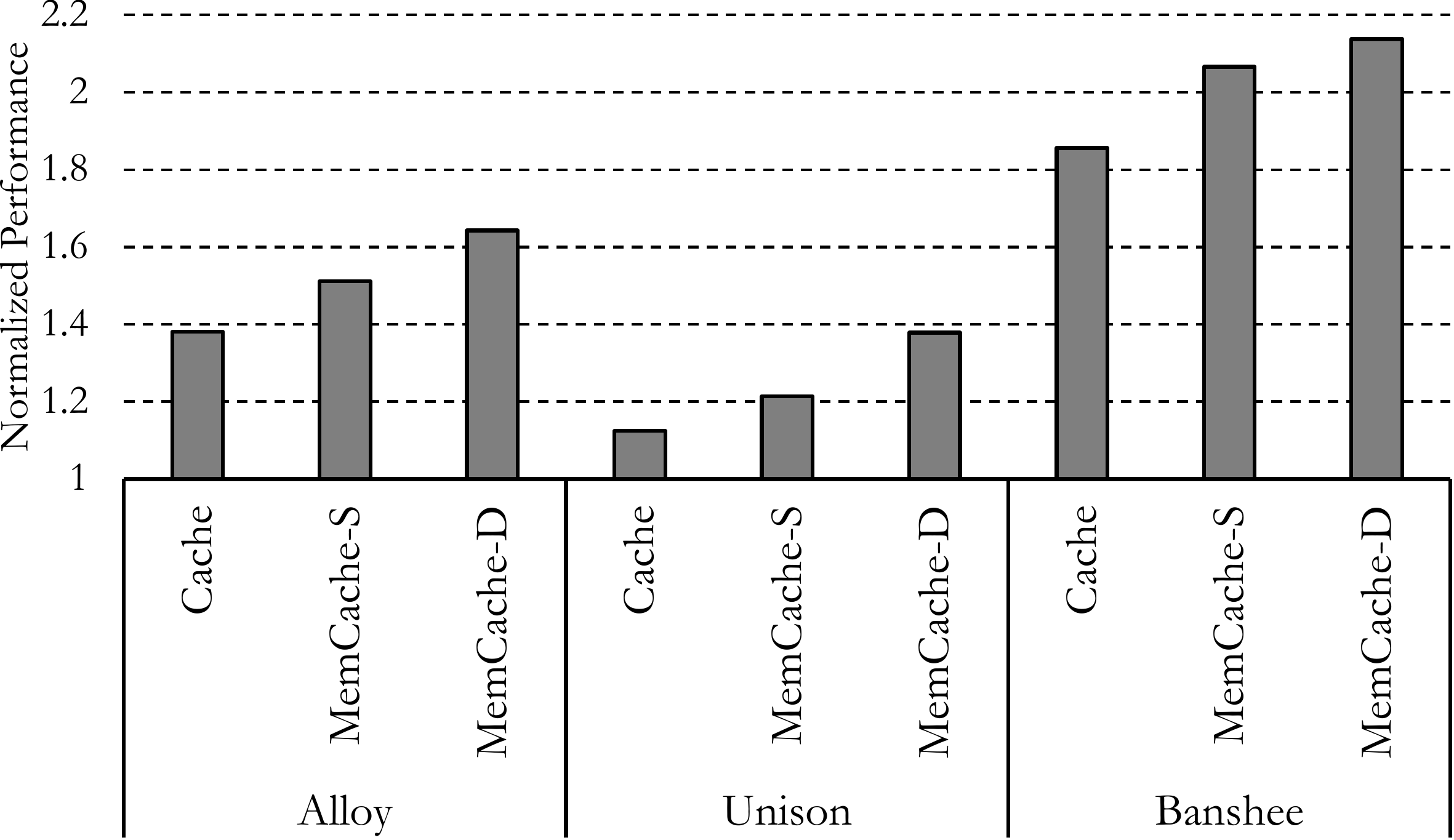}
 \caption{The sensitivity of performance to the cache organization of \methodname{MemCache}. Performance is normalized to a baseline without die-stacked DRAM.
 \label{fig:memcache-cache-design}}
\end{figure}

The results reported for \methodname{MemCache} are obtained for a design where the cache part is organized as a \methodname{Banshee Cache} as it offers the highest performance. In this section, we evaluate other architectures for the cache portion of the die-stacked DRAM. Figure~\ref{fig:memcache-cache-design} shows the performance of \methodname{MemCache} with various cache designs, as compared to the performance of the corresponding cache design. \methodname{MemCache-S} with \methodname{Alloy Cache}/\methodname{Unison Cache}/\methodname{Banshee Cache} improves performance by 13\%/9\%/21\% over the corresponding cache design. Furthermore, \methodname{MemCache-D} with \methodname{Alloy Cache}/\methodname{Unison Cache}/\methodname{Banshee Cache} enhances the performance of the corresponding cache design by 26\%/25\%/28\%. The performance improvement of \methodname{MemCache-S/D} with \methodname{Alloy Cache} as compared to the full-cache design is 13\%/26\% on average and up to 45\%/103\%. The performance improvement comes from: (1) 18\%/38\% lower MPKI, (2) 16\%/27\% less die-stacked DRAM bandwidth usage, and (3) 17\%/36\% lower traffic on the off-chip DRAM. \methodname{MemCache-S/D} with \methodname{Unison Cache} improves performance by 9\%/25\% on average and up to 66\%/174\% over the full-cache desgin. The performance enhancement comes from: (1) 8\%/49\% MPKI reduction, (2) 18\%/35\% less bandwidth pressure on the die-stacked DRAM, and (3) 5\%/33\% lower off-chip DRAM traffic.

The results indicate that \methodname{MemCache} is not limited to a specific cache design and can be used with other cache organizations, as well. Moreover, even with \methodname{MemCache}, the performance is significantly affected by the cache architecture. Since most of the bottlenecks on the road to achieve the peak performance are  cache-induced (e.g., tag-checking bandwidth and limited associativity), and not memory-induced, the architecture of the cache portion has a decisive role in the overall performance. 

\subsection{How Long are the Pages Hot?}
\label{eval:how_long}

\begin{figure}[t]
 \centering
 \includegraphics[width=.48\textwidth]{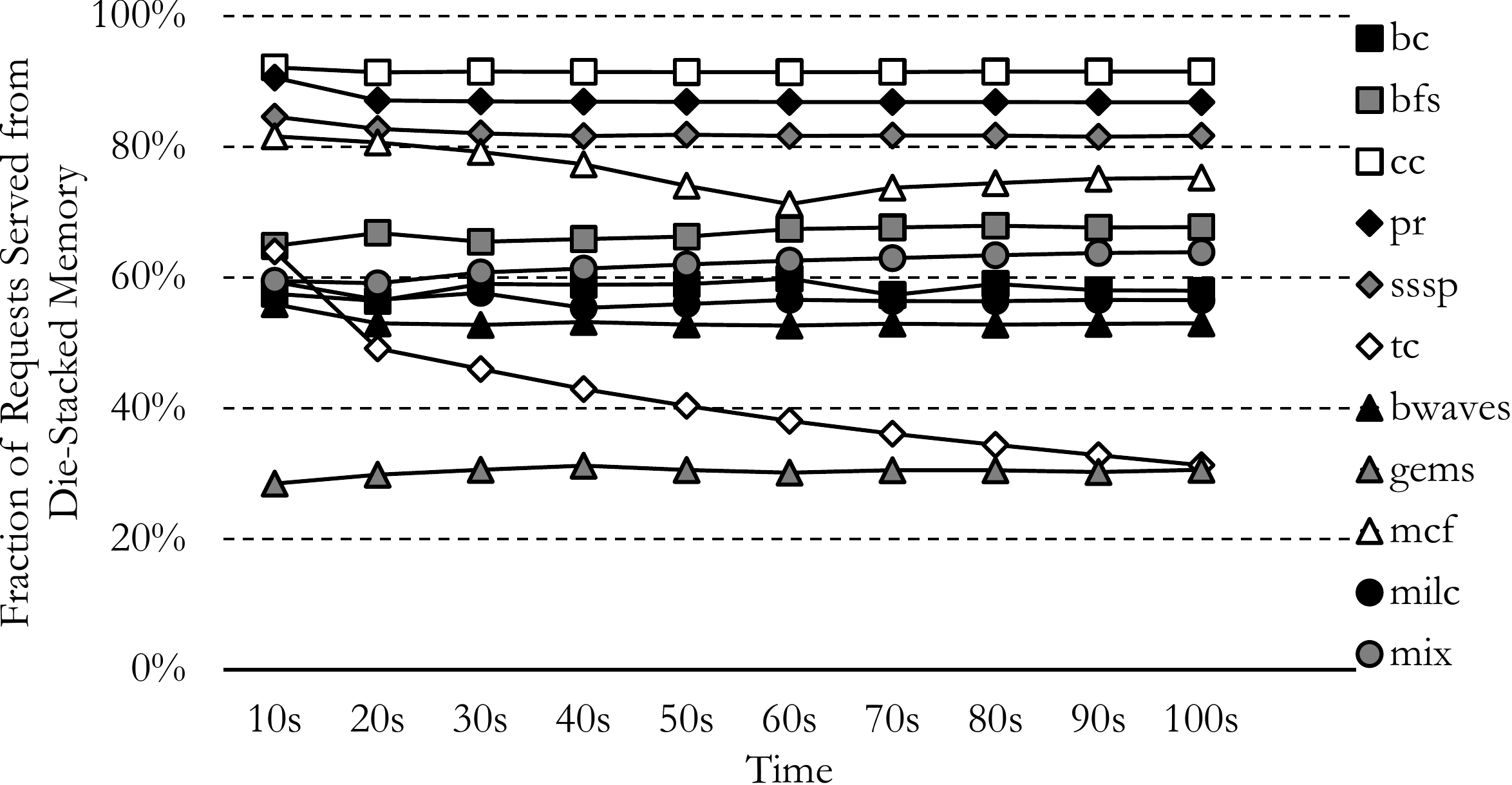}
 \caption{Fraction of requests that are served from the memory part of MemCache over time.
 \label{fig:hit-rate-time}}
\end{figure}

We fill the memory part of \methodname{MemCache} with hot pages that are distinguished at compile time. Any static and profile-based approach has the risk that its results become stale over time. Here, we measure the \emph{fraction of the requests that are served from the memory part}, as time spent from the point where timing simulations have finished. Figure~\ref{fig:hit-rate-time} shows how this fraction changes over 100 seconds for 3~GB of the hottest pages (i.e., \methodname{MemCache-S}). As shown, except for one workload (i.e., \appname{tc}), the metric remains steady. The results indicate that: 

\begin{enumerate}
\item Hot pages stem fundamentally from the data access patterns of applications and stay hot throughout the whole execution. 

\item A static compile-time analysis with a representative input-set is able to easily detect hot pages in the context of a giga-scale die-stacked DRAM. 

\end{enumerate}

The \appname{tc} application exhibits significant dynamic data-dependent behavior (cf.~Figure~\ref{fig:mem-vs-cache}) that results in a sharp drop in the reuse of pages that are classified as hot at the beginning of the execution. One solution for such workloads is to dynamically re-identify and swap pages between die-stacked and off-chip memories, just like prior work~\cite{dong2010simple, meswani:heterogeneous, Sim:2014:THM:2742155.2742158}. Note that, even for such applications, as a part of the die-stacked DRAM in \methodname{MemCache} is managed as a cache, it quickly responds to the dynamic changes of applications. Therefore, there is room for significantly increasing the period of OS interventions (e.g., once every 100 seconds), in order to amortize the high latency/bandwidth overhead. Evaluating such ideas is beyond the reach of architectural simulators, and hence, we leave them for future work.

\subsection{Sensitivity to Inputset}
\label{eval:inputset}

To evaluate the impact of inputset on the effectiveness of \methodname{MemCache}, we consider two other inputsets (i.e., \componentname{Web}~\cite{davis2011university} and \componentname{Urand}~\cite{erdds1959random}) for \componentname{GAPBS} applications\footnote{We do not perform a similar experiment for \componentname{SPEC} programs since other standard inputsets of \componentname{SPEC} applications are unable to fill the capacity of the die-stacked DRAM in a reasonable simulation time.} and compare the results with those of the so-far--evaluated \componentname{Twitter} inputset~\cite{kwak2010twitter}.
Table~\ref{table:inputset} summarizes the key characteristics of applications when they run these inputsets. Corroborating the characterization study performed by Beamer et al.~\cite{beamer2015gap}, \componentname{Web} graphs, despite the large size, exhibit substantial locality owing to their topology and high average degree~\cite{davis2011university}. Synthetically-generated \componentname{Urand} graphs, on the other hand, manifest the worst-case locality as every vertex in the graph has equal probability to be a neighbor of every other vertex~\cite{erdds1959random}. \componentname{Twitter} graphs, as they come from real-world data, have characteristics that lie in between \componentname{Web} and \componentname{Urand}. 

\begin{table}[h]
\sffamily
\scriptsize
 \begin{center}
  \caption{LLSC MPKI and memory footprint of evaluated graph-processing applications when they use different inputsets.}
  \label{table:inputset}
    \resizebox{.48\textwidth}{!}{
        \renewcommand{\arraystretch}{1}
        \begin{tabular}{| c || c | c | c | c | c | c |}
         \hline
          \multirow{2}{*}{\bf Application} & \multicolumn{3}{c|}{\bf {LLSC MPKI}}  & \multicolumn{3}{c|}{\bf {Memory Footprint (GB)}} \\ \cline{2-7} 
         
             & \componentname{Twitter} & \componentname{Web} & \componentname{Urand} & \componentname{Twitter} & \componentname{Web} & \componentname{Urand} \\
         \hline
           {bc} & 61.4 & 10.2 & 71.3 & 92.9 & 107.9 & 304.4 \\
         \hline
           {bfs} & 32.9 & 8.0 & 58.4 & 113.3 & 68.6 & 234.8 \\
         \hline
           {cc} & 85.6 & 8.2 & 98.7 & 9.6 & 9.2 & 25.0 \\
         \hline
           {pr} & 129.9 & 19.2 & 179.1 & 76.5 & 216.9 & 102.6 \\
         \hline
           {sssp} & 73.4 & 23.8 & 98.2 & 140.1 & 94.9 & 496.0 \\
         \hline
           {tc} & 12.5 & 2.9 & 16.2 & 53.4 & 21.5 & 266.2 \\
         \hline
        \end{tabular}
    }
 \end{center}
\end{table}

\begin{figure}[t]
 \centering
 \includegraphics[width=.48\textwidth]{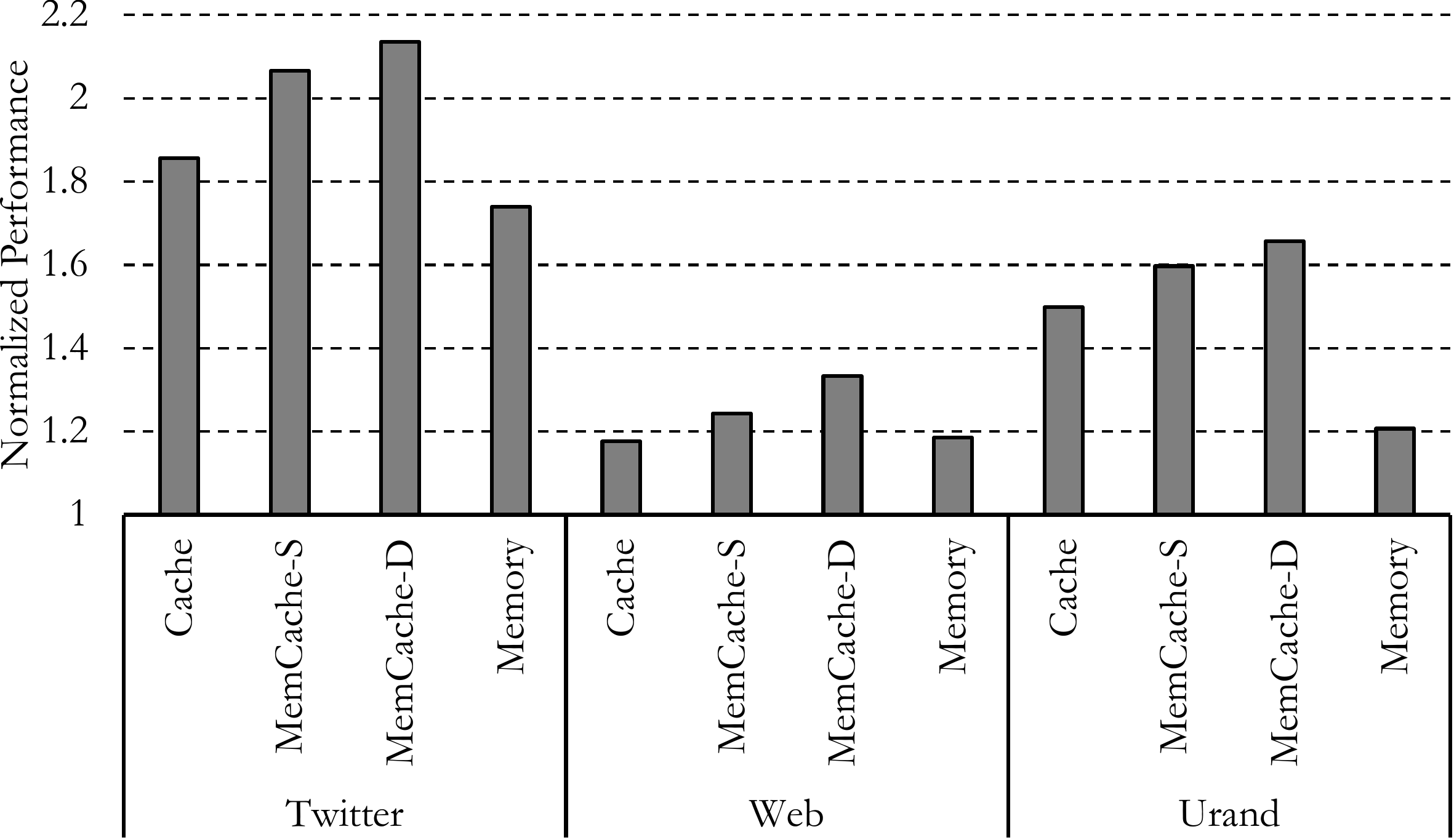}
 \caption{Sensitivity of performance improvement of various die-stacked DRAM organizations to inputset.
 \label{fig:inputset-sensitivity}}
\end{figure}

Figure~\ref{fig:inputset-sensitivity} shows the average performance improvement of die-stacked DRAM organizations when the applications use different inputsets. On average, with the \componentname{Twitter}/\componentname{Web}/\componentname{Urand} inputset, \methodname{MemCache-S} outperforms the best of cache and memory by 21\%/6\%/10\%. \methodname{MemCache-D} also outperforms \methodname{MemCache-S} by 7\%/9\%/6\%. While considerable, the performance improvement with \componentname{Web} and \componentname{Urand} inputsets is less than the improvement with the \componentname{Twitter} inputset. The improvement with \componentname{Web} is relatively low because most of the accesses hit in SRAM caches due to the high locality of graphs (cf.~Table~\ref{table:inputset}), reducing the bandwidth requirements of applications, and hence, downplaying the effect of bandwidth-improvement techniques. The lack of a considerable locality in \componentname{Urand} graphs, on the other face of the coin, defeat the caching and hot-page--detection policies used in the die-stacked DRAM organizations, giving rise to less performance improvement with these approaches. By and large, with all inputsets, \methodname{MemCache-S/D} consistently outperform both cache and memory designs, reinforcing our stance that using the die-stacked DRAM, partly as main memory and partly as cache, is the right design choice for modern throughput-oriented big-data applications.

\subsection{Sensitivity to Die-Stacked DRAM Size}
\label{eval:dram_size}

\begin{figure}[t]
 \centering
 \includegraphics[width=.48\textwidth]{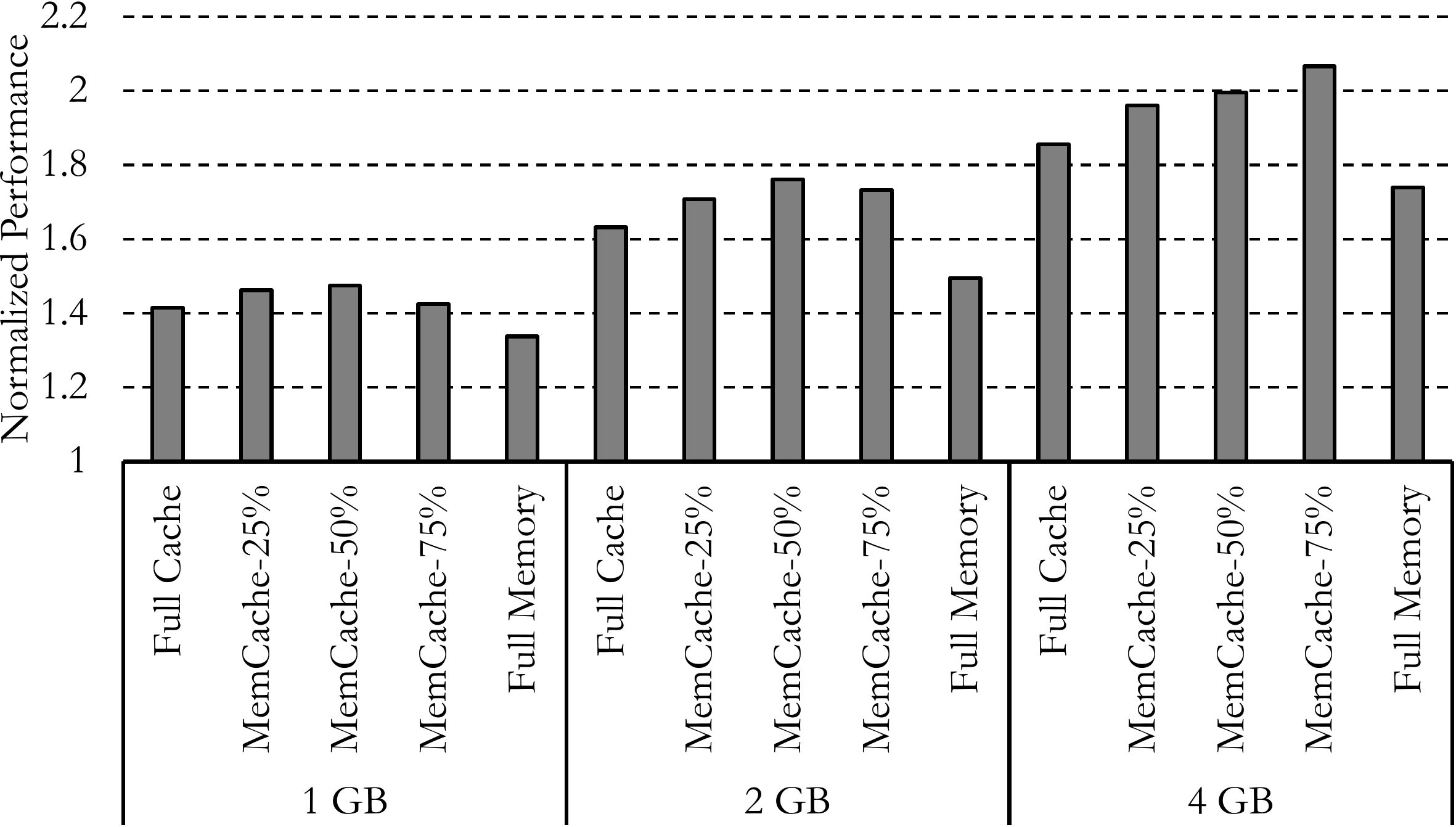}
 \caption{The sensitivity of performance improvement of various die-stacked DRAM organizations to the total capacity of die-stacked DRAM. \textit{MemCache-X\%} represents a hybrid design that dedicates \textit{X\%} of die-stacked DRAM capacity to memory and the rest to cache.
 \label{fig:dram_size}}
\end{figure}

Figure~\ref{fig:dram_size} shows the performance improvement of various methods with different die-stacked DRAM sizes. The figure shows that regardless of the die-stacked DRAM capacity, the hybrid design (i.e., \methodname{MemCache}) outperforms both cache and memory, because of \emph{simultaneously} reaping the benefits of both cache and memory. More to the point, with increasing the size of the die-stacked DRAM, the fraction of capacity that should be devoted to the memory in order to gain higher performance also increases. With increasing die-stacked DRAM capacity, the number of identified/allocated hot pages (cf.~Section~\ref{sec:motiv:what_frac}) and the accuracy of offline page classification (cf.~Section~\ref{sec:proposal:memory_portion}) increase, giving rise to increasing the memory fraction of optimal design. 

\section{Related Work}
\label{sec:related}

To the best of our knowledge, this is the first research that proposes to use the die-stacked DRAM partly as main memory and partly as a cache. Nevertheless, \methodname{MemCache}, in the sprite, is similar to \componentname{Hybrid} configuration of \componentname{Multi-Channel DRAM (MCDRAM)} modules in Intel's \componentname{Knights Landing} processor~\cite{sodani2016knightsieeemicro}. In \componentname{MCDRAM} modules of \componentname{Knights Landing} processor, similar to \methodname{MemCache-S}, a fixed portion of the die-stacked DRAM is devoted to memory, and the rest acts as a cache. The \emph{programmer} is responsible for allocating data structures in the memory portion of \componentname{MCDRAM} modules, using high-level instructions like {\ttfamily {hbw$\_$malloc}} and {\ttfamily FASTMEM}. In this paper, we showed that such a heavy burden on the programmer could be lifted with compile-time page classification. We showed that the large capacity of the die-stacked DRAM paves the way for offloading the hot-pages--identification task to the compiler, providing programming ease (cf.~Section~\ref{sec:proposal:memory_portion}). Moreover, we signified the potentials for dynamism in partitioning the capacity of the die-stacked DRAM, showing that the dynamic partitioning is able to outperform the static one by as much as 84\% (cf.~Section~\ref{sec:eval:perf}).

Some pieces of prior work~\cite{Agarwal:2015:PPS:2694344.2694381, loh2012challenges, meswani2014toward} also proposed to profile applications and manage the die-stacked DRAM based on the outcome. However, all of these approaches use the die-stacked DRAM entirely as a part of main memory and neither proposed nor discussed a design that uses the die-stacked DRAM partly as main memory and partly as a cache.

Chou et al.~\cite{Chou:2014:CTM:2742155.2742157} proposed \methodname{Cameo} to minimize the number of accesses to the backend storage (e.g., SSD or Disk). \methodname{Cameo} manages die-stacked and off-chip DRAM in different address spaces to increase the physical address space of the system while attempting to preserve the benefits of caching via keeping recently-accessed data in the high-bandwidth DRAM. \methodname{MemCache}, however, uses a part of the die-stacked DRAM as the main memory to enable efficient access to hot data structures without the overhead of tag-checking. \methodname{MemCache} is distinct from \methodname{Cameo} but brings most of its benefits for free. Yet, \methodname{Cameo} is orthogonal to \methodname{MemCache} since the cache part of the die-stacked DRAM in \methodname{MemCache} can be managed like \methodname{Cameo}. 

\methodname{Tertiary Caching} was proposed in the context of multi-socket systems. In \methodname{Tertiary Caching}~\cite{thekkath1997evaluation, zhang1997reducing}, a portion of each node's local memory is managed as a cache for caching only data objects that are allocated in remote nodes. While there are similarities between \methodname{MemCache} and \methodname{Tertiary Caching}, they are conceptually different. In \methodname{MemCache}, the memory part is used for keeping only hot pages, while in \methodname{Tertiary Caching}, memory hosts all local pages. Besides, \methodname{MemCache} uses the cache for the transient datasets of applications, while \methodname{Tertiary Caching} dedicates a part of memory to the cache in order to cache only remotely-allocated data objects.

In addition to schemes thoroughly discussed in Section~\ref{sec:background}, there are other proposals for managing the die-stacked DRAM as a cache. Loh and Hill~\cite{loh:efficiently} suggested a set-associative DRAM cache in which the tags and data of each set fit in a single DRAM row, allowing it to serve cache hits without the need to open more than one DRAM row. Jiang et al.~\cite{jiang2010chop} proposed \methodname{Chop}, a design that caches only pages with high expected reuse to avoid the significant bandwidth overhead of page-based caches. Loh~\cite{Loh:2009:EED:1669112.1669139} proposed organizing each DRAM cache set as a multi-level queue to evict dead-on-arrival cache lines immediately after insertion. Gulur et al.~\cite{Gulur:2014:BDC:2742155.2742160} proposed \methodname{Bi-Modal Dram Cache} to obtain the advantages of both block-based and page-based caches via dynamically choosing the caching granularity. Sim et al.~\cite{sim2012mostly} proposed techniques that enable accessing off-chip DRAM when the die-stacked DRAM cache bandwidth is highly utilized. In this paper, we showed that a large DRAM cache is suboptimal to a design that uses both cache and memory at the same time.

Numerous strategies were proposed to improve the efficiency of die-stacked DRAM. \methodname{Dap}~\cite{gaur2017near} and \methodname{Batman}~\cite{chou2017batman} attempt to maximize the total bandwidth utilization of systems with both die-stacked and DIMM-based DRAM modules. These approaches forbid data movement from the off-chip DRAM to the die-stacked DRAM when the bandwidth usage of the die-stacked DRAM exceeds a certain threshold to efficiently exploit the available bandwidth of both DRAM components. Franey and Lipasti~\cite{franey2015tag} proposed \methodname{Tag Tables}, a technique for compressing the tag array of DRAM caches to enable fabricating it in SRAM. \methodname{Accord}~\cite{youngaccord} provides associativity for direct-mapped \methodname{Alloy Cache} using way-prediction and way-install. Chou et al.~\cite{baer-isca} suggested \methodname{Bear} for reducing the bandwidth pressure of die-stacked DRAM caches. \methodname{Bear} decreases the tag-checking bandwidth by maintaining certain status details of the DRAM cache in the active die. Moreover, it samples replacement decisions, effectively trading hit ratio for bandwidth efficiency. Young et al.~\cite{young2017dice} introduced \methodname{DICE}, a technique for compressing DRAM caches mainly for reducing bandwidth usage and then for having a higher effective capacity. \methodname{DICE} attempts to get multiple useful blocks in a single DRAM access by compressing data and adaptively adjusting the set-indexing scheme of the cache. Huang and Nagarajan~\cite{huang2014atcache} proposed \methodname{ATCache} to provide faster access to the tag array of a DRAM cache. \methodname{ATCache} exploits the empirically-observed spatio-temporal locality in tag accesses and caches/prefetches some tags of the die-stacked DRAM in an SRAM structure. \methodname{Candy}~\cite{chou2016candy} and \methodname{C$^3$D}~\cite{huang2016c} facilitate the use of DRAM caches in the context of multi-node systems by making them coherent. Most of these approaches are orthogonal to our work and can be used together.

\section{Conclusion}

Die-stacked DRAM has shown the potential to break the bandwidth wall. The research community has evaluated two extreme use cases of the die-stacked DRAM: (1) as a sizable memory-side cache, and (2) as a part of software-visible main memory. In this work, we showed that both designs are suboptimal to a scheme that uses the die-stacked DRAM partly as main memory and partly as a cache. By analyzing the access behavior of various big-data applications, we observed that there are many hot pages with a significant number of accesses. We proposed \methodname{MemCache}, an approach that uses a portion of the die-stacked DRAM as the main memory for hosting hot pages and the rest as a cache for capturing the dynamic behavior of applications.

\bibliographystyle{IEEEtranS}
\bibliography{ref.bib}

\end{document}